%% file: nolta97.tex
\documentstyle[nolta]{article}
\newcommand{\alt}{\displaystyle{\mathop{<}_{\sim}}}

\begin{document}
\title{Wave Chaos in Quantum Pseudointegrable Billiards}
\author{
T. Shigehara$^{1}$, H. Mizoguchi$^{1}$, T. Mishima$^{1}$ 
and Taksu Cheon$^{2}$
} 
\affiliation{
$^{1}$Department of Information and Computer Sciences, Saitama University \\
Shimo-Okubo, Urawa, Saitama 338 Japan \\
e-mail: sigehara@ics.saitama-u.ac.jp \\
$^{2}$Laboratory of Physics, Kochi University of Technology \\
Tosa Yamada, Kochi 782 Japan \\
e-mail: cheon@mech.kochi-tech.ac.jp
}
\maketitle
\abstract
We clarify from a general perspective, 
the condition for the appearance of chaotic energy spectrum 
in quantum pseudointegrable billiards with a point scatterer 
inside. 
\endabstract

\section{Introduction} 

The quantum billiard with point scatterers is a quasi-exactly solvable 
model which is closely related to real systems. 
The billiards are a natural idealization 
of the particle motion in bounded systems. 
The single-electron problem in mesoscopic structures 
is now a possible setting, owing to the rapid progress 
in the mesoscopic technology. 
Real systems are, however, not free from impurities which affect the particle 
motion inside. In the presence of a small amount of contamination, even a 
single-electron problem becomes unmanageable. 
The modeling of the impurities with point scatterers makes the 
problem easy to handle without changing essential dynamics. 

A fundamental problem for the billiards considered here 
is to understand global behavior of the energy spectrum in the parameter 
space of particle energy and the strength of the scatterers. In particular, 
statistical properties of the spectrum are important because they reflect 
the degree of complexity of underlying dynamics. 
It is widely believed that integrable systems obeys Poisson statistics, 
while the predictions of the Gaussian Orthogonal Ensembles (GOE) describe 
chaotic systems \cite{BO89}. 

In this paper, we discuss the spectral properties of quantum billiards 
with a single point scatterer. It should be noticed that the 
nature of classical motion in such systems is pseudointegrable \cite{RB81} 
in the sense 
that unstable trajectories, which hit the point scatterer, are of measure 
zero. As a result, it is expected that the energy spectrum of a quantum 
analogue does not substantially differ from Poisson statistics. However, 
as shown in this paper, quantization induces the chaotic spectrum under a 
certain condition. 
This phenomenon might be called {\em wave chaos} \cite{AS91} 
because its origin is the wavelike nature of quantum dynamics. 
After deriving the eigenvalue equation 
with the help of self-adjoint extension theory in functional analysis, 
we discuss the statistical properties of the energy spectrum 
from a general perspective and deduce the condition for the appearance of 
chaotic spectrum in the quantum pseudointegrable system. 

\section{Formulation}

We start from an empty billiard in spatial dimension $d$, $d=1, 2, 3$. 
Let us consider a quantum point particle of mass $M$ moving 
freely in a bounded region $\Omega^{(d)}$. 
We impose the Dirichlet boundary condition such that wave functions 
vanish on the boundary of $\Omega^{(d)}$. 
The eigenvalues and the corresponding normalized eigenfunctions 
are denoted by 
$E_n$ and $\varphi_n(\vec{x})$ respectively; 
\begin{eqnarray}
\label{eq2-1}
H_0 \varphi _n({\vec x}) \equiv -{\nabla^2 \over {2M}} \varphi _n({\vec x})
= E _n \varphi _n({\vec x}). 
\end{eqnarray}
The Hamiltonian $H_0$ is the kinetic operator 
in $L^2(\Omega^{(d)})$ with domain 
$D(H_0)=H^2(\Omega^{(d)})\cap H^1_0(\Omega^{(d)})$ 
in terms of the Sobolev spaces. 
The Green's function of $H_0$ is given by 
\begin{eqnarray}
\label{eq2-2}
G^{(0)}({\vec x},{\vec x'};\omega ) = 
\sum\limits_{n=1}^\infty 
{{{\varphi _n({\vec x})\varphi _n({\vec x'})} 
\over {\omega -E_n}}}.
\end{eqnarray}
The average level density is given by 
\begin{eqnarray}
\label{eq2-3}
\rho^{(d)}_{av}(\omega) = \left\{ 
\begin{array}{ll}
\frac{M^{1/2}\Omega^{(1)}}{2^{1/2}\pi}\frac{1}{\sqrt{\omega}}, & 
\ \ \ d=1, \\ 
\frac{M\Omega^{(2)}}{2\pi}, & 
\ \ \ d=2, \\ 
\frac{M^{3/2}\Omega^{(3)}}{2^{1/2}\pi^2}\sqrt{\omega}, & 
\ \ \ d=3,  
\end{array} \right. 
\end{eqnarray}
where we denote the ``volume'' of $\Omega^{(d)}$ by the same symbol.  
(For example, the area of a two-dimensional region $\Omega^{(2)}$ is 
simply denoted by $\Omega^{(2)}$.)  

We now place a single point scatterer at 
${\vec x_0}$ inside the billiard. 
Naively, one defines the scatterer in terms of the $d$-dimensional 
Dirac's delta function;  
\begin{eqnarray}
\label{eq2-4}
H=H_0+ v \delta^{(d)} ({\vec x}-{\vec x_0}).
\end{eqnarray}
However, the Hamiltonian $H$ is not mathematically sound for $d \geq 2$. 
It is easy to see that 
the eigenvalue equation of $H$ is reduced to 
\begin{eqnarray}
\label{eq2-5}
\sum_{n=1}^{\infty}
\frac{\varphi_n(\vec{x}_0)^2}{\omega-E_n}=v^{-1}.  
\end{eqnarray}
Keeping Eq.(\ref{eq2-3}) in mind, we realize that 
the infinite series in Eq.(\ref{eq2-5}) does not converge for 
$d \geq 2$, since the average of the numerator (among many $n$) 
is energy-independent;  
$\left< \varphi_n (\vec{x}_0)^2 \right> \simeq 1/\Omega^{(d)}$.   

To handle the divergence, a scheme for renormalization is called for.  
One of the most satisfying schemes is given by the self-adjoint extension 
theory of functional analysis \cite{AG88}. 
We first consider in $L^2(\Omega^{(d)})$ the nonnegative operator 
\begin{eqnarray}
\label{eq2-7}
H_{\vec{x}_0} = - \left. {\nabla^2 \over {2M}} 
\right|_{C^{\infty}_0 (\Omega^{(d)}-\vec{x}_0)}  
\end{eqnarray}
with its closure ${\bar H_{\vec{x}_0}}$ in $L^2(\Omega^{(d)})$.   
Namely, we restrict $D(H_0)$ to the functions 
which vanish at the location of the point scatterer. 
By using integration by parts, 
it is easy to prove that 
the operator ${\bar H_{\vec{x}_0}}$ is symmetric (Hermitian). 
But it is not self-adjoint. Indeed, the equation 
\begin{eqnarray}
\label{eq2-8}
{\bar H_{\vec{x}_0}}^* \psi_{\omega}({\vec x}) = 
\omega \psi_{\omega}({\vec x}), \ \ \ \psi \in D({\bar H_{\vec{x}_0}}^*)  
\end{eqnarray}
has a solution for $Im \ \omega \neq 0$ \cite{ZO80}; 
\begin{eqnarray}
\label{eq2-9}
\psi_{\omega}({\vec x}) =  
G^{(0)}({\vec x},{\vec x_0};\omega ), \ \ {\vec x} \in \Omega^{(d)}-\vec{x}_0 
\end{eqnarray}
indicating 
\begin{eqnarray}
\label{eq2-10}
D(\bar{H}_{\vec{x}_0}^*) & 
\! \! \! \! \! 
= & 
\! \! \! \! \! 
D(\bar{H}_{\vec{x}_0}) \oplus N(\bar{H}_{\vec{x}_0}^* -\omega) 
\oplus N(\bar{H}_{\vec{x}_0}^* -\bar{\omega}) \nonumber \\
& 
\! \! \! \! \! 
\neq & 
\! \! \! \! \! 
D(\bar{H}_{\vec{x}_0}),  
\end{eqnarray}
where $N(A)$ is the kernel of an operator $A$. 
Since ${\bar H_{\vec{x}_0}}$ has the deficiency indices $(1,1)$, 
${\bar H_{\vec{x}_0}}$ has one-parameter family of self-adjoint extensions 
$H_{\theta}$ \cite{RS75}; 
\begin{eqnarray}
\label{eq2-11}
D(H_{\theta}) & 
\! \! \! \! \! 
= & 
\! \! \! \! \! 
\{ f \vert 
f = \varphi + c  ( \psi_{i\Lambda} 
- e^{i\theta} \psi_{-i\Lambda} ); \nonumber \\ 
&& \varphi\in D(\bar{H}_{\vec{x}_0}), c\in {\bf C}, 
0 \leq \theta < 2\pi \}, \nonumber \\ 
H_{\theta} f & 
\! \! \! \! \! 
= &
\! \! \! \! \! 
\bar{H}_{\vec{x}_0} \varphi+i \Lambda c 
( \psi_{i\Lambda} + e^{i\theta} \psi_{-i\Lambda} ),   
\end{eqnarray}
where $\Lambda >0$ is an arbitrary mass scale. 
With the aid of Krein's formula, 
we can write down 
the Green's function for the Hamiltonian $H_{\theta}$ as 
\begin{eqnarray}
\label{eq2-12}
G_{\theta}(\vec{x},\vec{x}';\omega) & 
\! \! \! \! \! 
= & 
\! \! \! \! \!   
G^{(0)}(\vec{x},\vec{x}';\omega) \nonumber \\
& 
\! \! \! \! \! \! \! \! \! \! \! \! \! \! \! \! \! \! \! \! \! \! 
+ & 
\! \! \! \! \! \! \! \! \! \! \! \! \! \! 
G^{(0)}(\vec{x},\vec{x}_{0};\omega)
T_{\theta}(\omega)
G^{(0)}(\vec{x}_{0},\vec{x}';\omega). 
\end{eqnarray}
The transition matrix (T-matrix) 
$T_{\theta}$ is calculated by 
\begin{eqnarray}
\label{eq2-13}
T_{\theta}(\omega) =\frac{1-e^{i\theta}}
{(\omega-i\Lambda) c_{i\Lambda}(\omega)-
e^{i\theta}(\omega+i\Lambda) c_{-i\Lambda}(\omega)}, 
\end{eqnarray}
where
\begin{eqnarray}
\label{eq2-14}
c_{\pm i\Lambda}(\omega)=
\int_{\Omega^{(d)}}
\! \! \! \! \!   
G^{(0)}(\vec{x},\vec{x}_{0};\omega)
G^{(0)}(\vec{x},\vec{x}_{0};\pm i\Lambda)d\vec{x}. 
\end{eqnarray}
Using the resolvent equation, we have 
\begin{eqnarray}
\label{eq2-15}
T_{\theta}(\omega)=(v_{\theta}^{-1}-G(\omega))^{-1}, 
\end{eqnarray}
where 
\begin{eqnarray}
\label{eq2-16}
v_{\theta}^{-1}=\Lambda \cot \frac{\theta}{2}
\sum_{n=1}^{\infty}
\frac{\varphi_{n}(\vec{x}_{0})^{2}}{E_{n}^2+\Lambda^{2}}, 
\end{eqnarray}
\vspace*{-6mm}
\begin{eqnarray}
\label{eq2-17}
G(\omega)=\sum_{n=1}^{\infty}
\varphi_{n}(\vec{x}_{0})^{2}
(\frac{1}{\omega-E_{n}}+\frac{E_{n}} {E_{n}^2+\Lambda^{2}}).
\end{eqnarray}
The constant $v_{\theta}$ is a coupling constant of the point scatterer, 
the value of which ranges over the whole real number   
as one varies $0\leq\theta <2\pi$.  
It follows from Eq.(\ref{eq2-15}) that 
the eigenvalues of $H_{\theta}$ are determined by 
\begin{eqnarray}
\label{eq2-18}
G(\omega)=v_{\theta}^{-1}.
\end{eqnarray}
On any interval $(E_n,E_{n+1})$, 
$G$ is a monotonically decreasing function of $\omega$ that ranges 
over the whole real number. 
This means that Eq.(\ref{eq2-18}) has a single solution on 
each interval for any $v_{\theta}$. 
The eigenfunction of $H_{\theta}$ corresponding to an eigenvalue $\omega_n$ 
is given by 
\begin{eqnarray}
\label{eq2-19}
\psi_n (\vec{x}) \propto G^{(0)}(\vec{x},\vec{x}_0;\omega_n). 
\end{eqnarray}

\section{Condition for Strong Coupling}

\subsection{$d=1$}

In one dimension, 
each infinite series on RHS in Eq.(\ref{eq2-17}) converges separately. 
Thus, if we set 
\begin{eqnarray}
\label{eq3a-1}
v^{-1}=v_{\theta}^{-1}-
\sum_{n=1}^{\infty}
\varphi_{n}(\vec{x}_{0})^{2}\frac{E_{n}} {E_{n}^2+\Lambda^{2}}, 
\end{eqnarray}
Eq.(\ref{eq2-18}) is reduced to Eq.(\ref{eq2-5});  
A somewhat complicated argument above leads us to 
the well-known result with Dirac's delta. 
In the following, we consider Eq.(\ref{eq2-5}), 
the LHS of which is denoted by $G^{(1)}(\omega)$.   

To obtain each solution of Eq.(\ref{eq2-5}), 
a numerical task is needed in general. 
However, since our main purpose lies 
in the statistical properties of spectrum,  
we proceed further by introducing some approximations 
without losing the essence, while still keeping loss 
of generality minimal \cite{SC97}.  
The first is that the value of   
$\varphi_n (\vec{x}_0)^2$ is replaced by 
its average among many $n$;  
\footnote{
With some minor modifications if necessary, 
a major part of the argument here 
is applicable to higher dimensions. 
We thus explicitly signify ``$d=1$'' only 
for the assertions specific to one dimension.}
\begin{eqnarray}
\label{eq3a-2}
\varphi_n (\vec{x}_0)^2 \simeq 
\langle \varphi_n (\vec{x}_0)^2 \rangle
=1/\Omega^{(d)}. 
\end{eqnarray}
Since the statistics 
is taken within a large number of, sometimes thousands of   
eigenstates, Eq.(\ref{eq3a-2}) is quite satisfactory. 
Keeping $\left| \varphi_n(\vec{x}_0) \right| 
\simeq 1/\sqrt{\Omega^{(d)}}$ in mind, 
we recognize from Eq.(\ref{eq2-19}) with Eq.(\ref{eq2-2}) that 
if $\omega_m \simeq E_m$ (or $E_{m+1}$) for some $m$, 
then $\psi_m \simeq \varphi_m$ (or $\varphi_{m+1}$),  
implying that only $\psi_m$ with an eigenvalue 
$\omega_m \simeq (E_{m}+E_{m+1})/2$ is distorted 
by a point scatterer. 
For such $\omega_m$, since 
the contributions on the summation of $G^{(d)}$ from 
the terms with $n \simeq m$ cancel each other,  
$G^{(d)}$ can be estimated by a principal integral;   
\begin{eqnarray}
\label{eq3a-3}
G^{(d)}(\omega_m) \simeq g^{(d)}(\omega_m), \ \ \ 
\omega_m \simeq \frac{E_{m}+E_{m+1}}{2}, 
\end{eqnarray} 
\vspace*{-6mm}
\begin{eqnarray}
\label{eq3a-4}
g^{(d)}(\omega) =  
\langle \varphi_{n}(\vec{x}_{0})^{2} \rangle  
P  \int_{0}^{\infty} 
\frac{\rho_{av}^{(d)}(E)}{\omega-E}dE, & d=1,  
\end{eqnarray} 
where we have defined a continuous function $g^{(d)}$ of $\omega$ 
which behaves like an interpolation of the inflection 
points of $G^{(d)}$. 
Inserting Eq.(\ref{eq2-3}) into Eq.(\ref{eq3a-4}) and 
using the elementary indefinite integral 
\begin{eqnarray}
\label{eq3a-5}
\int \frac{dE}{(\omega-E)\sqrt{E}} = 
\frac{1}{\sqrt{\omega}}
\ln \left| \frac{\sqrt{\omega}+\sqrt{E}}{\sqrt{\omega}-\sqrt{E}} \right| 
\end{eqnarray}
for $\omega>0$, we obtain for $\omega_m \simeq (E_{m}+E_{m+1})/2$  
\begin{eqnarray}
\label{eq3a-6}
G^{(d)}(\omega_m) \simeq 0, & d=1,   
\end{eqnarray} 
indicating that 
the maximal coupling of a point scatterer 
is attained when the strength $v$ satisfies  
\begin{eqnarray}
\label{eq3a-7}
v^{-1} \simeq 0, & d=1,   
\end{eqnarray} 
which is going well with our intuition. 

The ``width'' of the strong coupling region 
(allowable error of $v^{-1}$ in Eq.(\ref{eq3a-7}))
is estimated by considering a linearized eigenvalue equation. 
Expanding $G^{(d)}$ at 
$\omega_m \simeq (E_{m}+E_{m+1})/2$, 
we can rewrite 
the eigenvalue equation as 
\begin{eqnarray}
\label{eq3a-8}
G^{(d)'}(\omega_m)(\omega-\omega_m) \simeq v^{-1}-G^{(d)}(\omega_m). 
\end{eqnarray} 
In order that Eq.(\ref{eq3a-8}) has a solution 
$\omega\simeq \omega_m$, the range of RHS has to be restricted to 
\begin{eqnarray}
\label{eq3a-9}
\left| v^{-1} - G^{(d)}(\omega_m) \right| \ \ \alt \ \
\frac{\Delta^{(d)}(\omega_m)}{2},  
\end{eqnarray} 
\vspace*{-4mm}
\begin{eqnarray}
\label{eq3a-10}
\Delta^{(d)}(\omega_m) \equiv   
\left| G^{(d)'}(\omega_m) \right| 
\rho_{av}^{(d)}(\omega_m)^{-1}, 
\end{eqnarray} 
where we have defined the width $\Delta^{(d)}$ which 
is nothing but the average variance of the linearized $G^{(d)}$ 
on the interval $(E_m,E_{m+1})$. 
Using the approximation (\ref{eq3a-2}), 
the value of $\left| G^{(d)'}(\omega_m) \right|$ 
can be estimated as follows;  
\begin{eqnarray}
\label{eq3a-11} 
\left| G^{(d)'}(\omega_m) \right| 
& = & \sum_{n=1}^{\infty} \left(
\frac{\varphi_{n}(\vec{x}_{0})}
{\omega_m-E_{n}} \right)^2 \nonumber \\
& \simeq & \langle \varphi_{n}(\vec{x}_{0})^{2} \rangle 
\sum_{n=1}^{\infty} 
\frac{2}
{\{(n-\frac{1}{2})\rho_{av}^{(d)}(\omega_m)^{-1} \}^2} \nonumber \\
& = &   
8 \langle \varphi_{n}(\vec{x}_{0})^{2} \rangle 
\rho_{av}^{(d)}(\omega_m)^2 
\sum_{n=1}^{\infty} \frac{1}{(2n-1)^2} \nonumber \\
& = & 
\pi^{2} \langle \varphi_{n}(\vec{x}_{0})^{2} \rangle 
\rho_{av}^{(d)}(\omega_m)^2. 
\end{eqnarray}
The second equality follows from the approximation that 
the unperturbed eigenvalues are distributed with a mean 
interval $\rho_{av}^{(d)}(\omega_m)^{-1}$ in the whole energy region. 
This assumption is quite satisfactory, 
since the denominator of $G^{(d)'}(\omega)$ is of the order of 
$(\omega-E_n)^2$, indicating that the summation in 
Eq.(\ref{eq3a-11}) converges rapidly. 
From Eqs.(\ref{eq2-3}) and (\ref{eq3a-2}) for $d=1$, we obtain 
\begin{eqnarray}
\label{eq3a-12}
\Delta^{(1)}(\omega_m) \simeq 
\frac{\pi M^{1/2}}{2^{1/2}}\frac{1}{\sqrt{\omega_m}}, 
\end{eqnarray} 
which is inversely proportional to square root of the energy $\omega$. 
We can summarize the finding in one dimension as follows; 
The effect of a point scatterer of coupling strength $v$ 
is substantial only in the eigenstates with eigenvalue $\omega$ such that 
\begin{eqnarray}
\label{eq3a-13}
\left| v^{-1} \right| \alt \frac{\Delta^{(1)}(\omega)}{2} \simeq  
\frac{\pi M^{1/2}}{2^{3/2}}\frac{1}{\sqrt{\omega}}.   
\end{eqnarray}

\subsection{$d=2, 3$} 

It is easy to apply the argument above to higher dimensions.  
Since each series on RHS in Eq.(\ref{eq2-17}) diverges  
when summed separately for $d=2,3$, 
we have to resort to Eq.(\ref{eq2-18}).  
As a result, Eq.(\ref{eq3a-4}) is replaced by 
\begin{eqnarray}
\label{eq3b-1}
g^{(d)}(\omega) =  
\langle \varphi_{n}(\vec{x}_{0})^{2} \rangle  
P  \int_{0}^{\infty} 
\left( \frac{1}{\omega-E}+\frac{E}{E^{2}+\Lambda^2} \right) \nonumber
\end{eqnarray} 
\vspace*{-4mm}
\begin{eqnarray}
\times \rho_{av}^{(d)} (E)dE, & d=2,3. 
\end{eqnarray} 
Using the elementary indefinite integrals, 
\begin{eqnarray}
\label{eq3b-2}
\int \left( \frac{1}{\omega-E}+\frac{E}{E^{2}+\Lambda^2} \right) dE  = 
-\ln \frac{ | \omega-E | }
{\sqrt{ E^2 + \Lambda^2}}, 
\end{eqnarray}
\begin{eqnarray}
\int \left( \frac{1}{\omega-E}+\frac{E}{E^{2}+\Lambda^2} \right) \sqrt{E} dE 
= \sqrt{\omega} \ln \left| \frac{\sqrt{\omega}+\sqrt{E}}
{\sqrt{\omega}-\sqrt{E}} \right| 
\nonumber 
\end{eqnarray}
\vspace*{-4mm}
\begin{eqnarray}
- \frac{1}{2}\sqrt{\frac{\Lambda}{2}} 
\ln \left( \frac{E+\sqrt{2\Lambda E}+\Lambda}{E-\sqrt{2\Lambda E}+\Lambda} 
\right) 
\nonumber 
\end{eqnarray}
\vspace*{-4mm}
\begin{eqnarray}
- \sqrt{\frac{\Lambda}{2}}
\left\{ \arctan(\sqrt{\frac{2E}{\Lambda}}+1)+
\arctan(\sqrt{\frac{2E}{\Lambda}}-1) \right\}, 
\nonumber 
\end{eqnarray}
\vspace*{-4mm}
\begin{eqnarray}
\label{eq3b-3}
{\rm for \ \ } \omega>0, 
\end{eqnarray}
we obtain for $\omega_m \simeq (E_m + E_{m+1})/2$
\begin{eqnarray}
\label{eq3b-4}
G^{(d)}(\omega_m) \simeq \left\{ 
\begin{array}{ll}
\frac{M}{2\pi} \ln \frac{\omega_m}{\Lambda}, & d=2, \\
-\frac{M^{3/2}\Lambda^{1/2}}{2\pi}, & d=3. 
\end{array} \right. 
\end{eqnarray} 
The width $\Delta^{(d)}$ is estimated by Eq.(\ref{eq3a-10}) 
with Eq.(\ref{eq3a-11}). 
Using Eq.(\ref{eq2-3}), we find 
\begin{eqnarray}
\label{eq3b-5}
\Delta^{(d)}(\omega_m) \simeq \left\{ 
\begin{array}{ll}
\frac{\pi M}{2}, & d=2, \\
\frac{M^{3/2}}{2^{1/2}}\sqrt{\omega_m}, & d=3. 
\end{array} \right. 
\end{eqnarray}
From Eq.(\ref{eq3a-9}) with $v$ replaced by $v_{\theta}$, 
we realize that the strong coupling is attained 
under the condition 
\begin{eqnarray}
\label{eq3b-6}
\left| v_{\theta}^{-1} - \frac{M}{2\pi} \ln \frac{\omega}{\Lambda}  \right| 
\alt & 
\! \! \! \! \! \! \! \! \! \! 
\displaystyle \frac{\pi M}{4}, & d=2,  \\
\label{eq3b-7}
\left| v_{\theta}^{-1} + \frac{M^{3/2}\Lambda^{1/2}}{2\pi} \right| 
\alt & \displaystyle \frac{M^{3/2}}{2^{3/2}}\sqrt{\omega}, & d=3. 
\end{eqnarray}

\section{Numerical Example}

We consider a quantum particle with mass $M=1/2$ moving in 
a three-dimensional rectangular box  
with side-lengths $(l_{x},l_{y},l_{z})=(1.047, 1.186, 0.8049)$ 
and hence $\Omega^{(3)}=1$. The mass scale is set to $\Lambda=1$. 
A single point scatterer is placed at the center of the billiard. 
We take into account only the states with even parity in each direction, 
since the others are unaffected by the scatterer in this case. 
Fig.1 shows, for $v_{\theta}^{-1}=0, 10$ and $30$, 
the nearest-neighbor level spacing distribution $P(S)$, 
which is defined such that $P(S)dS$ is the probability to 
find the spacing between any two neighboring eigenstates in the 
interval $(S,S+dS)$. (The average spacing is normalized to one.) 
Generic integrable systems obey the Poisson distribution $P(S)=\exp (-S)$ 
(dotted line), while chaotic systems are described by the GOE prediction 
$P(S)=\pi S /2 \times \exp(-\pi S^2/4)$ (solid line). 
The statistics is taken within $\omega_{100} \sim \omega_{3100}$ 
in Fig.1 ($z_{1600}=8304$). 
According to Eq.(\ref{eq3b-7}),
the most strong coupling is attained for   
$v_{\theta}^{-1}=-M^{3/2}\Lambda^{1/2}/2\pi = -0.05626$. 
As $v_{\theta}^{-1}$ increases, 
$P(S)$ tends to approach the Poisson prediction. 
For $v_{\theta}^{-1} \simeq \Delta^{(3)}(\omega_{1600})/2 =11.3$, 
$P(S)$ is expected to be intermediate in shape between Poisson 
and GOE. 
The value can be considered as the upper bound of $v_{\theta}^{-1}$ 
for inducing a GOE-like shape in this energy region. 
With $v_{\theta}^{-1}$ beyond the bound, 
the system is not substantially different from the empty billiard, 
and as a result, $P(S)$ resembles to the Poisson distribution. 
These features are successfully reproduced in Fig.1. 
 
\section{Conclusion}

We have discussed the condition for the appearance of chaotic spectrum 
in the quantum pseudointegrable billiards with a single point scatterer 
inside. Chaotic spectrum appears if the conditions (\ref{eq3a-13}), 
(\ref{eq3b-6}) and (\ref{eq3b-7}) are satisfied for dimension $d=1,2,3$, 
respectively. 
In two dimension, 
the condition is described by a logarithmically energy-dependent strip with 
a constant width, indicating that the system recovers 
integrability in the high energy limit for any $v_{\theta}$. 
In three dimension, 
the spectrum shows GOE-like nature when $v_{\theta}^{-1}$ 
is within a band whose width increases parabolically as 
a function of the energy. This implies that the spectrum becomes chaotic 
at high energy for any $v_{\theta}$, which 
makes a clear contrast with two dimension.


\input{ps_lv0}

\input{ps_lv10}

\input{ps_lv30}

\begin{flushleft}
{\bf Fig.1.} 
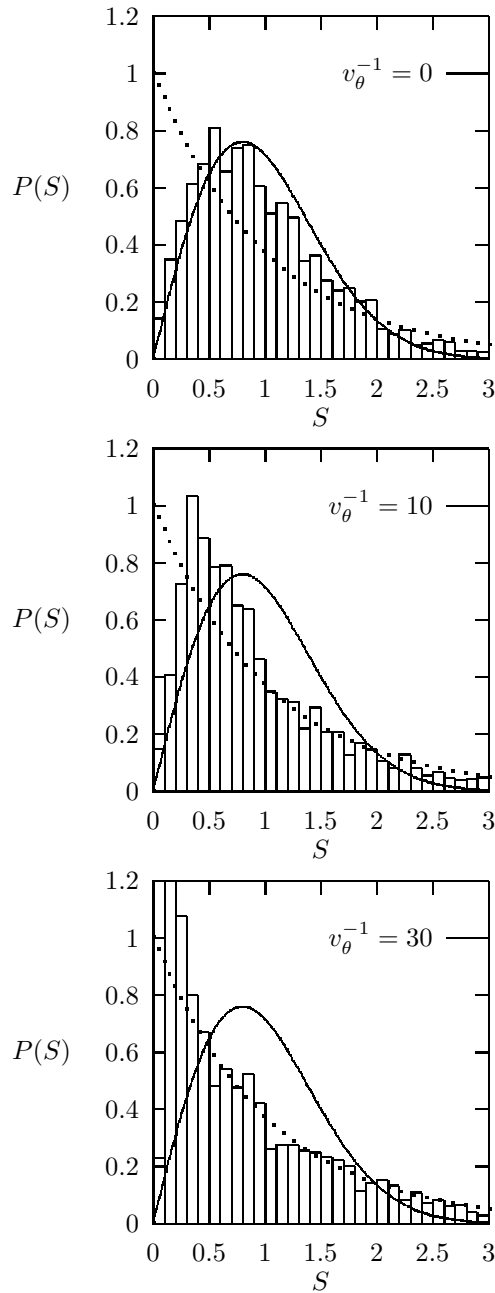
$P(S)$ for $v_{\theta}^{-1}=0, 10, 30$. 
The solid (dotted) line is GOE prediction (Poisson distribution). 
\end{flushleft}

\end{document}

%% file: ps_lv0.tex
\setlength{\unitlength}{0.240900pt}
\ifx\plotpoint\undefined\newsavebox{\plotpoint}\fi
\sbox{\plotpoint}{\rule[-0.200pt]{0.400pt}{0.400pt}}%
\begin{picture}(811,675)(0,0)
\font\gnuplot=cmr10 at 10pt
\gnuplot
\input{ps_inc}
\put(220.0,113.0){\rule[-0.200pt]{126.954pt}{0.400pt}}
\put(220.0,113.0){\rule[-0.200pt]{0.400pt}{129.845pt}}
\put(220.0,113.0){\rule[-0.200pt]{4.818pt}{0.400pt}}
\put(198,113){\makebox(0,0)[r]{0}}
\put(727.0,113.0){\rule[-0.200pt]{4.818pt}{0.400pt}}
\put(220.0,203.0){\rule[-0.200pt]{4.818pt}{0.400pt}}
\put(198,203){\makebox(0,0)[r]{0.2}}
\put(727.0,203.0){\rule[-0.200pt]{4.818pt}{0.400pt}}
\put(220.0,293.0){\rule[-0.200pt]{4.818pt}{0.400pt}}
\put(198,293){\makebox(0,0)[r]{0.4}}
\put(727.0,293.0){\rule[-0.200pt]{4.818pt}{0.400pt}}
\put(220.0,382.0){\rule[-0.200pt]{4.818pt}{0.400pt}}
\put(198,382){\makebox(0,0)[r]{0.6}}
\put(727.0,382.0){\rule[-0.200pt]{4.818pt}{0.400pt}}
\put(220.0,472.0){\rule[-0.200pt]{4.818pt}{0.400pt}}
\put(198,472){\makebox(0,0)[r]{0.8}}
\put(727.0,472.0){\rule[-0.200pt]{4.818pt}{0.400pt}}
\put(220.0,562.0){\rule[-0.200pt]{4.818pt}{0.400pt}}
\put(198,562){\makebox(0,0)[r]{1}}
\put(727.0,562.0){\rule[-0.200pt]{4.818pt}{0.400pt}}
\put(220.0,652.0){\rule[-0.200pt]{4.818pt}{0.400pt}}
\put(198,652){\makebox(0,0)[r]{1.2}}
\put(727.0,652.0){\rule[-0.200pt]{4.818pt}{0.400pt}}
\put(220.0,113.0){\rule[-0.200pt]{0.400pt}{4.818pt}}
\put(220,68){\makebox(0,0){0}}
\put(220.0,632.0){\rule[-0.200pt]{0.400pt}{4.818pt}}
\put(308.0,113.0){\rule[-0.200pt]{0.400pt}{4.818pt}}
\put(308,68){\makebox(0,0){0.5}}
\put(308.0,632.0){\rule[-0.200pt]{0.400pt}{4.818pt}}
\put(396.0,113.0){\rule[-0.200pt]{0.400pt}{4.818pt}}
\put(396,68){\makebox(0,0){1}}
\put(396.0,632.0){\rule[-0.200pt]{0.400pt}{4.818pt}}
\put(483.0,113.0){\rule[-0.200pt]{0.400pt}{4.818pt}}
\put(483,68){\makebox(0,0){1.5}}
\put(483.0,632.0){\rule[-0.200pt]{0.400pt}{4.818pt}}
\put(571.0,113.0){\rule[-0.200pt]{0.400pt}{4.818pt}}
\put(571,68){\makebox(0,0){2}}
\put(571.0,632.0){\rule[-0.200pt]{0.400pt}{4.818pt}}
\put(659.0,113.0){\rule[-0.200pt]{0.400pt}{4.818pt}}
\put(659,68){\makebox(0,0){2.5}}
\put(659.0,632.0){\rule[-0.200pt]{0.400pt}{4.818pt}}
\put(747.0,113.0){\rule[-0.200pt]{0.400pt}{4.818pt}}
\put(747,68){\makebox(0,0){3}}
\put(747.0,632.0){\rule[-0.200pt]{0.400pt}{4.818pt}}
\put(220.0,113.0){\rule[-0.200pt]{126.954pt}{0.400pt}}
\put(747.0,113.0){\rule[-0.200pt]{0.400pt}{129.845pt}}
\put(220.0,652.0){\rule[-0.200pt]{126.954pt}{0.400pt}}
\put(45,382){\makebox(0,0){$P(S)$}}
\put(483,23){\makebox(0,0){$S$}}
\put(220.0,113.0){\rule[-0.200pt]{0.400pt}{129.845pt}}
\put(220,562){\usebox{\plotpoint}}
\put(220.00,562.00){\usebox{\plotpoint}}
\put(227.86,542.80){\usebox{\plotpoint}}
\multiput(231,536)(7.983,-19.159){0}{\usebox{\plotpoint}}
\put(236.10,523.75){\usebox{\plotpoint}}
\put(243.93,504.54){\usebox{\plotpoint}}
\multiput(246,500)(9.282,-18.564){0}{\usebox{\plotpoint}}
\put(252.97,485.86){\usebox{\plotpoint}}
\put(261.56,466.97){\usebox{\plotpoint}}
\multiput(262,466)(9.282,-18.564){0}{\usebox{\plotpoint}}
\put(271.38,448.70){\usebox{\plotpoint}}
\multiput(273,446)(9.282,-18.564){0}{\usebox{\plotpoint}}
\put(280.87,430.25){\usebox{\plotpoint}}
\multiput(283,426)(11.513,-17.270){0}{\usebox{\plotpoint}}
\put(291.52,412.47){\usebox{\plotpoint}}
\multiput(294,408)(10.080,-18.144){0}{\usebox{\plotpoint}}
\put(301.83,394.47){\usebox{\plotpoint}}
\multiput(304,391)(12.453,-16.604){0}{\usebox{\plotpoint}}
\put(313.53,377.35){\usebox{\plotpoint}}
\multiput(315,375)(11.000,-17.601){0}{\usebox{\plotpoint}}
\put(324.97,360.04){\usebox{\plotpoint}}
\multiput(325,360)(12.453,-16.604){0}{\usebox{\plotpoint}}
\multiput(331,352)(12.064,-16.889){0}{\usebox{\plotpoint}}
\put(337.22,343.29){\usebox{\plotpoint}}
\multiput(341,338)(13.287,-15.945){0}{\usebox{\plotpoint}}
\put(350.20,327.10){\usebox{\plotpoint}}
\multiput(352,325)(13.287,-15.945){0}{\usebox{\plotpoint}}
\multiput(357,319)(13.287,-15.945){0}{\usebox{\plotpoint}}
\put(363.67,311.33){\usebox{\plotpoint}}
\multiput(368,307)(13.287,-15.945){0}{\usebox{\plotpoint}}
\put(377.83,296.17){\usebox{\plotpoint}}
\multiput(378,296)(13.287,-15.945){0}{\usebox{\plotpoint}}
\multiput(383,290)(15.945,-13.287){0}{\usebox{\plotpoint}}
\put(392.46,281.54){\usebox{\plotpoint}}
\multiput(394,280)(14.676,-14.676){0}{\usebox{\plotpoint}}
\multiput(399,275)(14.676,-14.676){0}{\usebox{\plotpoint}}
\put(407.69,267.54){\usebox{\plotpoint}}
\multiput(410,266)(14.676,-14.676){0}{\usebox{\plotpoint}}
\multiput(415,261)(16.207,-12.966){0}{\usebox{\plotpoint}}
\put(423.46,254.12){\usebox{\plotpoint}}
\multiput(426,252)(16.207,-12.966){0}{\usebox{\plotpoint}}
\multiput(431,248)(16.207,-12.966){0}{\usebox{\plotpoint}}
\put(439.62,241.10){\usebox{\plotpoint}}
\multiput(441,240)(18.564,-9.282){0}{\usebox{\plotpoint}}
\multiput(447,237)(16.207,-12.966){0}{\usebox{\plotpoint}}
\put(456.59,229.33){\usebox{\plotpoint}}
\multiput(457,229)(17.798,-10.679){0}{\usebox{\plotpoint}}
\multiput(462,226)(18.564,-9.282){0}{\usebox{\plotpoint}}
\multiput(468,223)(16.207,-12.966){0}{\usebox{\plotpoint}}
\put(474.11,218.34){\usebox{\plotpoint}}
\multiput(478,216)(17.798,-10.679){0}{\usebox{\plotpoint}}
\multiput(483,213)(18.564,-9.282){0}{\usebox{\plotpoint}}
\put(492.15,208.11){\usebox{\plotpoint}}
\multiput(494,207)(19.271,-7.708){0}{\usebox{\plotpoint}}
\multiput(499,205)(18.564,-9.282){0}{\usebox{\plotpoint}}
\multiput(505,202)(17.798,-10.679){0}{\usebox{\plotpoint}}
\put(510.63,198.75){\usebox{\plotpoint}}
\multiput(515,197)(17.798,-10.679){0}{\usebox{\plotpoint}}
\multiput(520,194)(19.690,-6.563){0}{\usebox{\plotpoint}}
\put(529.61,190.56){\usebox{\plotpoint}}
\multiput(531,190)(17.798,-10.679){0}{\usebox{\plotpoint}}
\multiput(536,187)(19.271,-7.708){0}{\usebox{\plotpoint}}
\multiput(541,185)(19.690,-6.563){0}{\usebox{\plotpoint}}
\put(548.60,182.36){\usebox{\plotpoint}}
\multiput(552,181)(19.271,-7.708){0}{\usebox{\plotpoint}}
\multiput(557,179)(19.690,-6.563){0}{\usebox{\plotpoint}}
\put(568.00,175.00){\usebox{\plotpoint}}
\multiput(568,175)(19.271,-7.708){0}{\usebox{\plotpoint}}
\multiput(573,173)(19.271,-7.708){0}{\usebox{\plotpoint}}
\multiput(578,171)(20.473,-3.412){0}{\usebox{\plotpoint}}
\put(587.62,168.55){\usebox{\plotpoint}}
\multiput(589,168)(19.271,-7.708){0}{\usebox{\plotpoint}}
\multiput(594,166)(20.352,-4.070){0}{\usebox{\plotpoint}}
\multiput(599,165)(19.690,-6.563){0}{\usebox{\plotpoint}}
\put(607.41,162.52){\usebox{\plotpoint}}
\multiput(610,162)(19.271,-7.708){0}{\usebox{\plotpoint}}
\multiput(615,160)(20.473,-3.412){0}{\usebox{\plotpoint}}
\multiput(621,159)(20.352,-4.070){0}{\usebox{\plotpoint}}
\put(627.44,157.43){\usebox{\plotpoint}}
\multiput(631,156)(20.352,-4.070){0}{\usebox{\plotpoint}}
\multiput(636,155)(20.473,-3.412){0}{\usebox{\plotpoint}}
\multiput(642,154)(20.352,-4.070){0}{\usebox{\plotpoint}}
\put(647.59,152.76){\usebox{\plotpoint}}
\multiput(652,151)(20.352,-4.070){0}{\usebox{\plotpoint}}
\multiput(657,150)(20.473,-3.412){0}{\usebox{\plotpoint}}
\put(667.73,148.05){\usebox{\plotpoint}}
\multiput(668,148)(20.352,-4.070){0}{\usebox{\plotpoint}}
\multiput(673,147)(20.352,-4.070){0}{\usebox{\plotpoint}}
\multiput(678,146)(20.473,-3.412){0}{\usebox{\plotpoint}}
\put(688.12,144.18){\usebox{\plotpoint}}
\multiput(689,144)(20.352,-4.070){0}{\usebox{\plotpoint}}
\multiput(694,143)(20.473,-3.412){0}{\usebox{\plotpoint}}
\multiput(700,142)(20.352,-4.070){0}{\usebox{\plotpoint}}
\put(708.58,141.00){\usebox{\plotpoint}}
\multiput(710,141)(20.352,-4.070){0}{\usebox{\plotpoint}}
\multiput(715,140)(20.473,-3.412){0}{\usebox{\plotpoint}}
\multiput(721,139)(20.352,-4.070){0}{\usebox{\plotpoint}}
\put(728.99,137.40){\usebox{\plotpoint}}
\multiput(731,137)(20.755,0.000){0}{\usebox{\plotpoint}}
\multiput(736,137)(20.473,-3.412){0}{\usebox{\plotpoint}}
\multiput(742,136)(20.352,-4.070){0}{\usebox{\plotpoint}}
\put(747,135){\usebox{\plotpoint}}
\put(220,113){\usebox{\plotpoint}}
\multiput(220.59,113.00)(0.477,2.269){7}{\rule{0.115pt}{1.780pt}}
\multiput(219.17,113.00)(5.000,17.306){2}{\rule{0.400pt}{0.890pt}}
\multiput(225.59,134.00)(0.482,1.847){9}{\rule{0.116pt}{1.500pt}}
\multiput(224.17,134.00)(6.000,17.887){2}{\rule{0.400pt}{0.750pt}}
\multiput(231.59,155.00)(0.477,2.269){7}{\rule{0.115pt}{1.780pt}}
\multiput(230.17,155.00)(5.000,17.306){2}{\rule{0.400pt}{0.890pt}}
\multiput(236.59,176.00)(0.477,2.269){7}{\rule{0.115pt}{1.780pt}}
\multiput(235.17,176.00)(5.000,17.306){2}{\rule{0.400pt}{0.890pt}}
\multiput(241.59,197.00)(0.477,2.157){7}{\rule{0.115pt}{1.700pt}}
\multiput(240.17,197.00)(5.000,16.472){2}{\rule{0.400pt}{0.850pt}}
\multiput(246.59,217.00)(0.482,1.756){9}{\rule{0.116pt}{1.433pt}}
\multiput(245.17,217.00)(6.000,17.025){2}{\rule{0.400pt}{0.717pt}}
\multiput(252.59,237.00)(0.477,2.046){7}{\rule{0.115pt}{1.620pt}}
\multiput(251.17,237.00)(5.000,15.638){2}{\rule{0.400pt}{0.810pt}}
\multiput(257.59,256.00)(0.477,2.046){7}{\rule{0.115pt}{1.620pt}}
\multiput(256.17,256.00)(5.000,15.638){2}{\rule{0.400pt}{0.810pt}}
\multiput(262.59,275.00)(0.477,1.935){7}{\rule{0.115pt}{1.540pt}}
\multiput(261.17,275.00)(5.000,14.804){2}{\rule{0.400pt}{0.770pt}}
\multiput(267.59,293.00)(0.482,1.485){9}{\rule{0.116pt}{1.233pt}}
\multiput(266.17,293.00)(6.000,14.440){2}{\rule{0.400pt}{0.617pt}}
\multiput(273.59,310.00)(0.477,1.823){7}{\rule{0.115pt}{1.460pt}}
\multiput(272.17,310.00)(5.000,13.970){2}{\rule{0.400pt}{0.730pt}}
\multiput(278.59,327.00)(0.477,1.601){7}{\rule{0.115pt}{1.300pt}}
\multiput(277.17,327.00)(5.000,12.302){2}{\rule{0.400pt}{0.650pt}}
\multiput(283.59,342.00)(0.482,1.304){9}{\rule{0.116pt}{1.100pt}}
\multiput(282.17,342.00)(6.000,12.717){2}{\rule{0.400pt}{0.550pt}}
\multiput(289.59,357.00)(0.477,1.489){7}{\rule{0.115pt}{1.220pt}}
\multiput(288.17,357.00)(5.000,11.468){2}{\rule{0.400pt}{0.610pt}}
\multiput(294.59,371.00)(0.477,1.378){7}{\rule{0.115pt}{1.140pt}}
\multiput(293.17,371.00)(5.000,10.634){2}{\rule{0.400pt}{0.570pt}}
\multiput(299.59,384.00)(0.477,1.267){7}{\rule{0.115pt}{1.060pt}}
\multiput(298.17,384.00)(5.000,9.800){2}{\rule{0.400pt}{0.530pt}}
\multiput(304.59,396.00)(0.482,0.852){9}{\rule{0.116pt}{0.767pt}}
\multiput(303.17,396.00)(6.000,8.409){2}{\rule{0.400pt}{0.383pt}}
\multiput(310.59,406.00)(0.477,1.044){7}{\rule{0.115pt}{0.900pt}}
\multiput(309.17,406.00)(5.000,8.132){2}{\rule{0.400pt}{0.450pt}}
\multiput(315.59,416.00)(0.477,0.933){7}{\rule{0.115pt}{0.820pt}}
\multiput(314.17,416.00)(5.000,7.298){2}{\rule{0.400pt}{0.410pt}}
\multiput(320.59,425.00)(0.477,0.710){7}{\rule{0.115pt}{0.660pt}}
\multiput(319.17,425.00)(5.000,5.630){2}{\rule{0.400pt}{0.330pt}}
\multiput(325.00,432.59)(0.491,0.482){9}{\rule{0.500pt}{0.116pt}}
\multiput(325.00,431.17)(4.962,6.000){2}{\rule{0.250pt}{0.400pt}}
\multiput(331.59,438.00)(0.477,0.599){7}{\rule{0.115pt}{0.580pt}}
\multiput(330.17,438.00)(5.000,4.796){2}{\rule{0.400pt}{0.290pt}}
\multiput(336.00,444.60)(0.627,0.468){5}{\rule{0.600pt}{0.113pt}}
\multiput(336.00,443.17)(3.755,4.000){2}{\rule{0.300pt}{0.400pt}}
\multiput(341.00,448.61)(0.909,0.447){3}{\rule{0.767pt}{0.108pt}}
\multiput(341.00,447.17)(3.409,3.000){2}{\rule{0.383pt}{0.400pt}}
\put(346,451.17){\rule{1.300pt}{0.400pt}}
\multiput(346.00,450.17)(3.302,2.000){2}{\rule{0.650pt}{0.400pt}}
\put(352,452.67){\rule{1.204pt}{0.400pt}}
\multiput(352.00,452.17)(2.500,1.000){2}{\rule{0.602pt}{0.400pt}}
\put(368,452.17){\rule{1.100pt}{0.400pt}}
\multiput(368.00,453.17)(2.717,-2.000){2}{\rule{0.550pt}{0.400pt}}
\multiput(373.00,450.95)(0.909,-0.447){3}{\rule{0.767pt}{0.108pt}}
\multiput(373.00,451.17)(3.409,-3.000){2}{\rule{0.383pt}{0.400pt}}
\multiput(378.00,447.95)(0.909,-0.447){3}{\rule{0.767pt}{0.108pt}}
\multiput(378.00,448.17)(3.409,-3.000){2}{\rule{0.383pt}{0.400pt}}
\multiput(383.00,444.93)(0.599,-0.477){7}{\rule{0.580pt}{0.115pt}}
\multiput(383.00,445.17)(4.796,-5.000){2}{\rule{0.290pt}{0.400pt}}
\multiput(389.00,439.93)(0.487,-0.477){7}{\rule{0.500pt}{0.115pt}}
\multiput(389.00,440.17)(3.962,-5.000){2}{\rule{0.250pt}{0.400pt}}
\multiput(394.00,434.93)(0.487,-0.477){7}{\rule{0.500pt}{0.115pt}}
\multiput(394.00,435.17)(3.962,-5.000){2}{\rule{0.250pt}{0.400pt}}
\multiput(399.59,428.59)(0.477,-0.599){7}{\rule{0.115pt}{0.580pt}}
\multiput(398.17,429.80)(5.000,-4.796){2}{\rule{0.400pt}{0.290pt}}
\multiput(404.59,422.65)(0.482,-0.581){9}{\rule{0.116pt}{0.567pt}}
\multiput(403.17,423.82)(6.000,-5.824){2}{\rule{0.400pt}{0.283pt}}
\multiput(410.59,415.26)(0.477,-0.710){7}{\rule{0.115pt}{0.660pt}}
\multiput(409.17,416.63)(5.000,-5.630){2}{\rule{0.400pt}{0.330pt}}
\multiput(415.59,407.93)(0.477,-0.821){7}{\rule{0.115pt}{0.740pt}}
\multiput(414.17,409.46)(5.000,-6.464){2}{\rule{0.400pt}{0.370pt}}
\multiput(420.59,400.37)(0.482,-0.671){9}{\rule{0.116pt}{0.633pt}}
\multiput(419.17,401.69)(6.000,-6.685){2}{\rule{0.400pt}{0.317pt}}
\multiput(426.59,391.60)(0.477,-0.933){7}{\rule{0.115pt}{0.820pt}}
\multiput(425.17,393.30)(5.000,-7.298){2}{\rule{0.400pt}{0.410pt}}
\multiput(431.59,382.60)(0.477,-0.933){7}{\rule{0.115pt}{0.820pt}}
\multiput(430.17,384.30)(5.000,-7.298){2}{\rule{0.400pt}{0.410pt}}
\multiput(436.59,373.60)(0.477,-0.933){7}{\rule{0.115pt}{0.820pt}}
\multiput(435.17,375.30)(5.000,-7.298){2}{\rule{0.400pt}{0.410pt}}
\multiput(441.59,365.09)(0.482,-0.762){9}{\rule{0.116pt}{0.700pt}}
\multiput(440.17,366.55)(6.000,-7.547){2}{\rule{0.400pt}{0.350pt}}
\multiput(447.59,355.60)(0.477,-0.933){7}{\rule{0.115pt}{0.820pt}}
\multiput(446.17,357.30)(5.000,-7.298){2}{\rule{0.400pt}{0.410pt}}
\multiput(452.59,346.60)(0.477,-0.933){7}{\rule{0.115pt}{0.820pt}}
\multiput(451.17,348.30)(5.000,-7.298){2}{\rule{0.400pt}{0.410pt}}
\multiput(457.59,337.26)(0.477,-1.044){7}{\rule{0.115pt}{0.900pt}}
\multiput(456.17,339.13)(5.000,-8.132){2}{\rule{0.400pt}{0.450pt}}
\multiput(462.59,328.09)(0.482,-0.762){9}{\rule{0.116pt}{0.700pt}}
\multiput(461.17,329.55)(6.000,-7.547){2}{\rule{0.400pt}{0.350pt}}
\multiput(468.59,318.26)(0.477,-1.044){7}{\rule{0.115pt}{0.900pt}}
\multiput(467.17,320.13)(5.000,-8.132){2}{\rule{0.400pt}{0.450pt}}
\multiput(473.59,308.60)(0.477,-0.933){7}{\rule{0.115pt}{0.820pt}}
\multiput(472.17,310.30)(5.000,-7.298){2}{\rule{0.400pt}{0.410pt}}
\multiput(478.59,299.60)(0.477,-0.933){7}{\rule{0.115pt}{0.820pt}}
\multiput(477.17,301.30)(5.000,-7.298){2}{\rule{0.400pt}{0.410pt}}
\multiput(483.59,291.09)(0.482,-0.762){9}{\rule{0.116pt}{0.700pt}}
\multiput(482.17,292.55)(6.000,-7.547){2}{\rule{0.400pt}{0.350pt}}
\multiput(489.59,281.60)(0.477,-0.933){7}{\rule{0.115pt}{0.820pt}}
\multiput(488.17,283.30)(5.000,-7.298){2}{\rule{0.400pt}{0.410pt}}
\multiput(494.59,272.60)(0.477,-0.933){7}{\rule{0.115pt}{0.820pt}}
\multiput(493.17,274.30)(5.000,-7.298){2}{\rule{0.400pt}{0.410pt}}
\multiput(499.59,264.37)(0.482,-0.671){9}{\rule{0.116pt}{0.633pt}}
\multiput(498.17,265.69)(6.000,-6.685){2}{\rule{0.400pt}{0.317pt}}
\multiput(505.59,255.60)(0.477,-0.933){7}{\rule{0.115pt}{0.820pt}}
\multiput(504.17,257.30)(5.000,-7.298){2}{\rule{0.400pt}{0.410pt}}
\multiput(510.59,246.93)(0.477,-0.821){7}{\rule{0.115pt}{0.740pt}}
\multiput(509.17,248.46)(5.000,-6.464){2}{\rule{0.400pt}{0.370pt}}
\multiput(515.59,238.93)(0.477,-0.821){7}{\rule{0.115pt}{0.740pt}}
\multiput(514.17,240.46)(5.000,-6.464){2}{\rule{0.400pt}{0.370pt}}
\multiput(520.59,231.65)(0.482,-0.581){9}{\rule{0.116pt}{0.567pt}}
\multiput(519.17,232.82)(6.000,-5.824){2}{\rule{0.400pt}{0.283pt}}
\multiput(526.59,224.26)(0.477,-0.710){7}{\rule{0.115pt}{0.660pt}}
\multiput(525.17,225.63)(5.000,-5.630){2}{\rule{0.400pt}{0.330pt}}
\multiput(531.59,217.26)(0.477,-0.710){7}{\rule{0.115pt}{0.660pt}}
\multiput(530.17,218.63)(5.000,-5.630){2}{\rule{0.400pt}{0.330pt}}
\multiput(536.59,210.26)(0.477,-0.710){7}{\rule{0.115pt}{0.660pt}}
\multiput(535.17,211.63)(5.000,-5.630){2}{\rule{0.400pt}{0.330pt}}
\multiput(541.00,204.93)(0.491,-0.482){9}{\rule{0.500pt}{0.116pt}}
\multiput(541.00,205.17)(4.962,-6.000){2}{\rule{0.250pt}{0.400pt}}
\multiput(547.59,197.59)(0.477,-0.599){7}{\rule{0.115pt}{0.580pt}}
\multiput(546.17,198.80)(5.000,-4.796){2}{\rule{0.400pt}{0.290pt}}
\multiput(552.59,191.59)(0.477,-0.599){7}{\rule{0.115pt}{0.580pt}}
\multiput(551.17,192.80)(5.000,-4.796){2}{\rule{0.400pt}{0.290pt}}
\multiput(557.00,186.93)(0.491,-0.482){9}{\rule{0.500pt}{0.116pt}}
\multiput(557.00,187.17)(4.962,-6.000){2}{\rule{0.250pt}{0.400pt}}
\multiput(563.00,180.93)(0.487,-0.477){7}{\rule{0.500pt}{0.115pt}}
\multiput(563.00,181.17)(3.962,-5.000){2}{\rule{0.250pt}{0.400pt}}
\multiput(568.00,175.93)(0.487,-0.477){7}{\rule{0.500pt}{0.115pt}}
\multiput(568.00,176.17)(3.962,-5.000){2}{\rule{0.250pt}{0.400pt}}
\multiput(573.00,170.94)(0.627,-0.468){5}{\rule{0.600pt}{0.113pt}}
\multiput(573.00,171.17)(3.755,-4.000){2}{\rule{0.300pt}{0.400pt}}
\multiput(578.00,166.93)(0.599,-0.477){7}{\rule{0.580pt}{0.115pt}}
\multiput(578.00,167.17)(4.796,-5.000){2}{\rule{0.290pt}{0.400pt}}
\multiput(584.00,161.94)(0.627,-0.468){5}{\rule{0.600pt}{0.113pt}}
\multiput(584.00,162.17)(3.755,-4.000){2}{\rule{0.300pt}{0.400pt}}
\multiput(589.00,157.95)(0.909,-0.447){3}{\rule{0.767pt}{0.108pt}}
\multiput(589.00,158.17)(3.409,-3.000){2}{\rule{0.383pt}{0.400pt}}
\multiput(594.00,154.94)(0.627,-0.468){5}{\rule{0.600pt}{0.113pt}}
\multiput(594.00,155.17)(3.755,-4.000){2}{\rule{0.300pt}{0.400pt}}
\multiput(599.00,150.95)(1.132,-0.447){3}{\rule{0.900pt}{0.108pt}}
\multiput(599.00,151.17)(4.132,-3.000){2}{\rule{0.450pt}{0.400pt}}
\multiput(605.00,147.95)(0.909,-0.447){3}{\rule{0.767pt}{0.108pt}}
\multiput(605.00,148.17)(3.409,-3.000){2}{\rule{0.383pt}{0.400pt}}
\multiput(610.00,144.95)(0.909,-0.447){3}{\rule{0.767pt}{0.108pt}}
\multiput(610.00,145.17)(3.409,-3.000){2}{\rule{0.383pt}{0.400pt}}
\multiput(615.00,141.95)(1.132,-0.447){3}{\rule{0.900pt}{0.108pt}}
\multiput(615.00,142.17)(4.132,-3.000){2}{\rule{0.450pt}{0.400pt}}
\put(621,138.17){\rule{1.100pt}{0.400pt}}
\multiput(621.00,139.17)(2.717,-2.000){2}{\rule{0.550pt}{0.400pt}}
\multiput(626.00,136.95)(0.909,-0.447){3}{\rule{0.767pt}{0.108pt}}
\multiput(626.00,137.17)(3.409,-3.000){2}{\rule{0.383pt}{0.400pt}}
\put(631,133.17){\rule{1.100pt}{0.400pt}}
\multiput(631.00,134.17)(2.717,-2.000){2}{\rule{0.550pt}{0.400pt}}
\put(636,131.17){\rule{1.300pt}{0.400pt}}
\multiput(636.00,132.17)(3.302,-2.000){2}{\rule{0.650pt}{0.400pt}}
\put(642,129.67){\rule{1.204pt}{0.400pt}}
\multiput(642.00,130.17)(2.500,-1.000){2}{\rule{0.602pt}{0.400pt}}
\put(647,128.17){\rule{1.100pt}{0.400pt}}
\multiput(647.00,129.17)(2.717,-2.000){2}{\rule{0.550pt}{0.400pt}}
\put(652,126.17){\rule{1.100pt}{0.400pt}}
\multiput(652.00,127.17)(2.717,-2.000){2}{\rule{0.550pt}{0.400pt}}
\put(657,124.67){\rule{1.445pt}{0.400pt}}
\multiput(657.00,125.17)(3.000,-1.000){2}{\rule{0.723pt}{0.400pt}}
\put(663,123.67){\rule{1.204pt}{0.400pt}}
\multiput(663.00,124.17)(2.500,-1.000){2}{\rule{0.602pt}{0.400pt}}
\put(668,122.67){\rule{1.204pt}{0.400pt}}
\multiput(668.00,123.17)(2.500,-1.000){2}{\rule{0.602pt}{0.400pt}}
\put(673,121.67){\rule{1.204pt}{0.400pt}}
\multiput(673.00,122.17)(2.500,-1.000){2}{\rule{0.602pt}{0.400pt}}
\put(678,120.67){\rule{1.445pt}{0.400pt}}
\multiput(678.00,121.17)(3.000,-1.000){2}{\rule{0.723pt}{0.400pt}}
\put(684,119.67){\rule{1.204pt}{0.400pt}}
\multiput(684.00,120.17)(2.500,-1.000){2}{\rule{0.602pt}{0.400pt}}
\put(689,118.67){\rule{1.204pt}{0.400pt}}
\multiput(689.00,119.17)(2.500,-1.000){2}{\rule{0.602pt}{0.400pt}}
\put(357.0,454.0){\rule[-0.200pt]{2.650pt}{0.400pt}}
\put(700,117.67){\rule{1.204pt}{0.400pt}}
\multiput(700.00,118.17)(2.500,-1.000){2}{\rule{0.602pt}{0.400pt}}
\put(705,116.67){\rule{1.204pt}{0.400pt}}
\multiput(705.00,117.17)(2.500,-1.000){2}{\rule{0.602pt}{0.400pt}}
\put(694.0,119.0){\rule[-0.200pt]{1.445pt}{0.400pt}}
\put(715,115.67){\rule{1.445pt}{0.400pt}}
\multiput(715.00,116.17)(3.000,-1.000){2}{\rule{0.723pt}{0.400pt}}
\put(710.0,117.0){\rule[-0.200pt]{1.204pt}{0.400pt}}
\put(731,114.67){\rule{1.204pt}{0.400pt}}
\multiput(731.00,115.17)(2.500,-1.000){2}{\rule{0.602pt}{0.400pt}}
\put(721.0,116.0){\rule[-0.200pt]{2.409pt}{0.400pt}}
\put(736.0,115.0){\rule[-0.200pt]{2.650pt}{0.400pt}}
\put(659,562){\makebox(0,0)[r]{$v_{\theta}^{-1}=0$}}
\put(681.0,562.0){\rule[-0.200pt]{15.899pt}{0.400pt}}
\put(220.0,113.0){\rule[-0.200pt]{0.400pt}{15.418pt}}
\put(220.0,177.0){\rule[-0.200pt]{4.336pt}{0.400pt}}
\put(238.0,113.0){\rule[-0.200pt]{0.400pt}{15.418pt}}
\put(220.0,113.0){\rule[-0.200pt]{4.336pt}{0.400pt}}
\put(238.0,113.0){\rule[-0.200pt]{0.400pt}{37.821pt}}
\put(238.0,270.0){\rule[-0.200pt]{4.095pt}{0.400pt}}
\put(255.0,113.0){\rule[-0.200pt]{0.400pt}{37.821pt}}
\put(238.0,113.0){\rule[-0.200pt]{4.095pt}{0.400pt}}
\put(255.0,113.0){\rule[-0.200pt]{0.400pt}{52.275pt}}
\put(255.0,330.0){\rule[-0.200pt]{4.336pt}{0.400pt}}
\put(273.0,113.0){\rule[-0.200pt]{0.400pt}{52.275pt}}
\put(255.0,113.0){\rule[-0.200pt]{4.336pt}{0.400pt}}
\put(273.0,113.0){\rule[-0.200pt]{0.400pt}{66.247pt}}
\put(273.0,388.0){\rule[-0.200pt]{4.095pt}{0.400pt}}
\put(290.0,113.0){\rule[-0.200pt]{0.400pt}{66.247pt}}
\put(273.0,113.0){\rule[-0.200pt]{4.095pt}{0.400pt}}
\put(290.0,113.0){\rule[-0.200pt]{0.400pt}{73.956pt}}
\put(290.0,420.0){\rule[-0.200pt]{4.336pt}{0.400pt}}
\put(308.0,113.0){\rule[-0.200pt]{0.400pt}{73.956pt}}
\put(290.0,113.0){\rule[-0.200pt]{4.336pt}{0.400pt}}
\put(308.0,113.0){\rule[-0.200pt]{0.400pt}{87.688pt}}
\put(308.0,477.0){\rule[-0.200pt]{4.095pt}{0.400pt}}
\put(325.0,113.0){\rule[-0.200pt]{0.400pt}{87.688pt}}
\put(308.0,113.0){\rule[-0.200pt]{4.095pt}{0.400pt}}
\put(325.0,113.0){\rule[-0.200pt]{0.400pt}{71.065pt}}
\put(325.0,408.0){\rule[-0.200pt]{4.336pt}{0.400pt}}
\put(343.0,113.0){\rule[-0.200pt]{0.400pt}{71.065pt}}
\put(325.0,113.0){\rule[-0.200pt]{4.336pt}{0.400pt}}
\put(343.0,113.0){\rule[-0.200pt]{0.400pt}{79.979pt}}
\put(343.0,445.0){\rule[-0.200pt]{4.336pt}{0.400pt}}
\put(361.0,113.0){\rule[-0.200pt]{0.400pt}{79.979pt}}
\put(343.0,113.0){\rule[-0.200pt]{4.336pt}{0.400pt}}
\put(361.0,113.0){\rule[-0.200pt]{0.400pt}{81.183pt}}
\put(361.0,450.0){\rule[-0.200pt]{4.095pt}{0.400pt}}
\put(378.0,113.0){\rule[-0.200pt]{0.400pt}{81.183pt}}
\put(361.0,113.0){\rule[-0.200pt]{4.095pt}{0.400pt}}
\put(378.0,113.0){\rule[-0.200pt]{0.400pt}{65.525pt}}
\put(378.0,385.0){\rule[-0.200pt]{4.336pt}{0.400pt}}
\put(396.0,113.0){\rule[-0.200pt]{0.400pt}{65.525pt}}
\put(378.0,113.0){\rule[-0.200pt]{4.336pt}{0.400pt}}
\put(396.0,113.0){\rule[-0.200pt]{0.400pt}{55.166pt}}
\put(396.0,342.0){\rule[-0.200pt]{4.095pt}{0.400pt}}
\put(413.0,113.0){\rule[-0.200pt]{0.400pt}{55.166pt}}
\put(396.0,113.0){\rule[-0.200pt]{4.095pt}{0.400pt}}
\put(413.0,113.0){\rule[-0.200pt]{0.400pt}{59.261pt}}
\put(413.0,359.0){\rule[-0.200pt]{4.336pt}{0.400pt}}
\put(431.0,113.0){\rule[-0.200pt]{0.400pt}{59.261pt}}
\put(413.0,113.0){\rule[-0.200pt]{4.336pt}{0.400pt}}
\put(431.0,113.0){\rule[-0.200pt]{0.400pt}{53.721pt}}
\put(431.0,336.0){\rule[-0.200pt]{4.095pt}{0.400pt}}
\put(448.0,113.0){\rule[-0.200pt]{0.400pt}{53.721pt}}
\put(431.0,113.0){\rule[-0.200pt]{4.095pt}{0.400pt}}
\put(448.0,113.0){\rule[-0.200pt]{0.400pt}{37.099pt}}
\put(448.0,267.0){\rule[-0.200pt]{4.336pt}{0.400pt}}
\put(466.0,113.0){\rule[-0.200pt]{0.400pt}{37.099pt}}
\put(448.0,113.0){\rule[-0.200pt]{4.336pt}{0.400pt}}
\put(466.0,113.0){\rule[-0.200pt]{0.400pt}{39.267pt}}
\put(466.0,276.0){\rule[-0.200pt]{4.095pt}{0.400pt}}
\put(483.0,113.0){\rule[-0.200pt]{0.400pt}{39.267pt}}
\put(466.0,113.0){\rule[-0.200pt]{4.095pt}{0.400pt}}
\put(483.0,113.0){\rule[-0.200pt]{0.400pt}{29.872pt}}
\put(483.0,237.0){\rule[-0.200pt]{4.336pt}{0.400pt}}
\put(501.0,113.0){\rule[-0.200pt]{0.400pt}{29.872pt}}
\put(483.0,113.0){\rule[-0.200pt]{4.336pt}{0.400pt}}
\put(501.0,113.0){\rule[-0.200pt]{0.400pt}{26.017pt}}
\put(501.0,221.0){\rule[-0.200pt]{4.336pt}{0.400pt}}
\put(519.0,113.0){\rule[-0.200pt]{0.400pt}{26.017pt}}
\put(501.0,113.0){\rule[-0.200pt]{4.336pt}{0.400pt}}
\put(519.0,113.0){\rule[-0.200pt]{0.400pt}{26.981pt}}
\put(519.0,225.0){\rule[-0.200pt]{4.095pt}{0.400pt}}
\put(536.0,113.0){\rule[-0.200pt]{0.400pt}{26.981pt}}
\put(519.0,113.0){\rule[-0.200pt]{4.095pt}{0.400pt}}
\put(536.0,113.0){\rule[-0.200pt]{0.400pt}{21.922pt}}
\put(536.0,204.0){\rule[-0.200pt]{4.336pt}{0.400pt}}
\put(554.0,113.0){\rule[-0.200pt]{0.400pt}{21.922pt}}
\put(536.0,113.0){\rule[-0.200pt]{4.336pt}{0.400pt}}
\put(554.0,113.0){\rule[-0.200pt]{0.400pt}{22.404pt}}
\put(554.0,206.0){\rule[-0.200pt]{4.095pt}{0.400pt}}
\put(571.0,113.0){\rule[-0.200pt]{0.400pt}{22.404pt}}
\put(554.0,113.0){\rule[-0.200pt]{4.095pt}{0.400pt}}
\put(571.0,113.0){\rule[-0.200pt]{0.400pt}{11.563pt}}
\put(571.0,161.0){\rule[-0.200pt]{4.336pt}{0.400pt}}
\put(589.0,113.0){\rule[-0.200pt]{0.400pt}{11.563pt}}
\put(571.0,113.0){\rule[-0.200pt]{4.336pt}{0.400pt}}
\put(589.0,113.0){\rule[-0.200pt]{0.400pt}{9.636pt}}
\put(589.0,153.0){\rule[-0.200pt]{4.095pt}{0.400pt}}
\put(606.0,113.0){\rule[-0.200pt]{0.400pt}{9.636pt}}
\put(589.0,113.0){\rule[-0.200pt]{4.095pt}{0.400pt}}
\put(606.0,113.0){\rule[-0.200pt]{0.400pt}{11.081pt}}
\put(606.0,159.0){\rule[-0.200pt]{4.336pt}{0.400pt}}
\put(624.0,113.0){\rule[-0.200pt]{0.400pt}{11.081pt}}
\put(606.0,113.0){\rule[-0.200pt]{4.336pt}{0.400pt}}
\put(624.0,113.0){\rule[-0.200pt]{0.400pt}{5.300pt}}
\put(624.0,135.0){\rule[-0.200pt]{4.336pt}{0.400pt}}
\put(642.0,113.0){\rule[-0.200pt]{0.400pt}{5.300pt}}
\put(624.0,113.0){\rule[-0.200pt]{4.336pt}{0.400pt}}
\put(642.0,113.0){\rule[-0.200pt]{0.400pt}{6.022pt}}
\put(642.0,138.0){\rule[-0.200pt]{4.095pt}{0.400pt}}
\put(659.0,113.0){\rule[-0.200pt]{0.400pt}{6.022pt}}
\put(642.0,113.0){\rule[-0.200pt]{4.095pt}{0.400pt}}
\put(659.0,113.0){\rule[-0.200pt]{0.400pt}{7.227pt}}
\put(659.0,143.0){\rule[-0.200pt]{4.336pt}{0.400pt}}
\put(677.0,113.0){\rule[-0.200pt]{0.400pt}{7.227pt}}
\put(659.0,113.0){\rule[-0.200pt]{4.336pt}{0.400pt}}
\put(677.0,113.0){\rule[-0.200pt]{0.400pt}{6.504pt}}
\put(677.0,140.0){\rule[-0.200pt]{4.095pt}{0.400pt}}
\put(694.0,113.0){\rule[-0.200pt]{0.400pt}{6.504pt}}
\put(677.0,113.0){\rule[-0.200pt]{4.095pt}{0.400pt}}
\put(694.0,113.0){\rule[-0.200pt]{0.400pt}{3.132pt}}
\put(694.0,126.0){\rule[-0.200pt]{4.336pt}{0.400pt}}
\put(712.0,113.0){\rule[-0.200pt]{0.400pt}{3.132pt}}
\put(694.0,113.0){\rule[-0.200pt]{4.336pt}{0.400pt}}
\put(712.0,113.0){\rule[-0.200pt]{0.400pt}{3.132pt}}
\put(712.0,126.0){\rule[-0.200pt]{4.095pt}{0.400pt}}
\put(729.0,113.0){\rule[-0.200pt]{0.400pt}{3.132pt}}
\put(712.0,113.0){\rule[-0.200pt]{4.095pt}{0.400pt}}
\put(729.0,113.0){\rule[-0.200pt]{0.400pt}{2.891pt}}
\put(729.0,125.0){\rule[-0.200pt]{4.336pt}{0.400pt}}
\put(747.0,113.0){\rule[-0.200pt]{0.400pt}{2.891pt}}
\put(729.0,113.0){\rule[-0.200pt]{4.336pt}{0.400pt}}
\end{picture}

%% file: ps_lv10.tex
\setlength{\unitlength}{0.240900pt}
\ifx\plotpoint\undefined\newsavebox{\plotpoint}\fi
\sbox{\plotpoint}{\rule[-0.200pt]{0.400pt}{0.400pt}}%
\begin{picture}(811,675)(0,0)
\font\gnuplot=cmr10 at 10pt
\gnuplot
\input{ps_inc}
\put(220.0,113.0){\rule[-0.200pt]{126.954pt}{0.400pt}}
\put(220.0,113.0){\rule[-0.200pt]{0.400pt}{129.845pt}}
\put(220.0,113.0){\rule[-0.200pt]{4.818pt}{0.400pt}}
\put(198,113){\makebox(0,0)[r]{0}}
\put(727.0,113.0){\rule[-0.200pt]{4.818pt}{0.400pt}}
\put(220.0,203.0){\rule[-0.200pt]{4.818pt}{0.400pt}}
\put(198,203){\makebox(0,0)[r]{0.2}}
\put(727.0,203.0){\rule[-0.200pt]{4.818pt}{0.400pt}}
\put(220.0,293.0){\rule[-0.200pt]{4.818pt}{0.400pt}}
\put(198,293){\makebox(0,0)[r]{0.4}}
\put(727.0,293.0){\rule[-0.200pt]{4.818pt}{0.400pt}}
\put(220.0,382.0){\rule[-0.200pt]{4.818pt}{0.400pt}}
\put(198,382){\makebox(0,0)[r]{0.6}}
\put(727.0,382.0){\rule[-0.200pt]{4.818pt}{0.400pt}}
\put(220.0,472.0){\rule[-0.200pt]{4.818pt}{0.400pt}}
\put(198,472){\makebox(0,0)[r]{0.8}}
\put(727.0,472.0){\rule[-0.200pt]{4.818pt}{0.400pt}}
\put(220.0,562.0){\rule[-0.200pt]{4.818pt}{0.400pt}}
\put(198,562){\makebox(0,0)[r]{1}}
\put(727.0,562.0){\rule[-0.200pt]{4.818pt}{0.400pt}}
\put(220.0,652.0){\rule[-0.200pt]{4.818pt}{0.400pt}}
\put(198,652){\makebox(0,0)[r]{1.2}}
\put(727.0,652.0){\rule[-0.200pt]{4.818pt}{0.400pt}}
\put(220.0,113.0){\rule[-0.200pt]{0.400pt}{4.818pt}}
\put(220,68){\makebox(0,0){0}}
\put(220.0,632.0){\rule[-0.200pt]{0.400pt}{4.818pt}}
\put(308.0,113.0){\rule[-0.200pt]{0.400pt}{4.818pt}}
\put(308,68){\makebox(0,0){0.5}}
\put(308.0,632.0){\rule[-0.200pt]{0.400pt}{4.818pt}}
\put(396.0,113.0){\rule[-0.200pt]{0.400pt}{4.818pt}}
\put(396,68){\makebox(0,0){1}}
\put(396.0,632.0){\rule[-0.200pt]{0.400pt}{4.818pt}}
\put(483.0,113.0){\rule[-0.200pt]{0.400pt}{4.818pt}}
\put(483,68){\makebox(0,0){1.5}}
\put(483.0,632.0){\rule[-0.200pt]{0.400pt}{4.818pt}}
\put(571.0,113.0){\rule[-0.200pt]{0.400pt}{4.818pt}}
\put(571,68){\makebox(0,0){2}}
\put(571.0,632.0){\rule[-0.200pt]{0.400pt}{4.818pt}}
\put(659.0,113.0){\rule[-0.200pt]{0.400pt}{4.818pt}}
\put(659,68){\makebox(0,0){2.5}}
\put(659.0,632.0){\rule[-0.200pt]{0.400pt}{4.818pt}}
\put(747.0,113.0){\rule[-0.200pt]{0.400pt}{4.818pt}}
\put(747,68){\makebox(0,0){3}}
\put(747.0,632.0){\rule[-0.200pt]{0.400pt}{4.818pt}}
\put(220.0,113.0){\rule[-0.200pt]{126.954pt}{0.400pt}}
\put(747.0,113.0){\rule[-0.200pt]{0.400pt}{129.845pt}}
\put(220.0,652.0){\rule[-0.200pt]{126.954pt}{0.400pt}}
\put(45,382){\makebox(0,0){$P(S)$}}
\put(483,23){\makebox(0,0){$S$}}
\put(220.0,113.0){\rule[-0.200pt]{0.400pt}{129.845pt}}
\put(220,562){\usebox{\plotpoint}}
\put(220.00,562.00){\usebox{\plotpoint}}
\put(227.86,542.80){\usebox{\plotpoint}}
\multiput(231,536)(7.983,-19.159){0}{\usebox{\plotpoint}}
\put(236.10,523.75){\usebox{\plotpoint}}
\put(243.93,504.54){\usebox{\plotpoint}}
\multiput(246,500)(9.282,-18.564){0}{\usebox{\plotpoint}}
\put(252.97,485.86){\usebox{\plotpoint}}
\put(261.56,466.97){\usebox{\plotpoint}}
\multiput(262,466)(9.282,-18.564){0}{\usebox{\plotpoint}}
\put(271.38,448.70){\usebox{\plotpoint}}
\multiput(273,446)(9.282,-18.564){0}{\usebox{\plotpoint}}
\put(280.87,430.25){\usebox{\plotpoint}}
\multiput(283,426)(11.513,-17.270){0}{\usebox{\plotpoint}}
\put(291.52,412.47){\usebox{\plotpoint}}
\multiput(294,408)(10.080,-18.144){0}{\usebox{\plotpoint}}
\put(301.83,394.47){\usebox{\plotpoint}}
\multiput(304,391)(12.453,-16.604){0}{\usebox{\plotpoint}}
\put(313.53,377.35){\usebox{\plotpoint}}
\multiput(315,375)(11.000,-17.601){0}{\usebox{\plotpoint}}
\put(324.97,360.04){\usebox{\plotpoint}}
\multiput(325,360)(12.453,-16.604){0}{\usebox{\plotpoint}}
\multiput(331,352)(12.064,-16.889){0}{\usebox{\plotpoint}}
\put(337.22,343.29){\usebox{\plotpoint}}
\multiput(341,338)(13.287,-15.945){0}{\usebox{\plotpoint}}
\put(350.20,327.10){\usebox{\plotpoint}}
\multiput(352,325)(13.287,-15.945){0}{\usebox{\plotpoint}}
\multiput(357,319)(13.287,-15.945){0}{\usebox{\plotpoint}}
\put(363.67,311.33){\usebox{\plotpoint}}
\multiput(368,307)(13.287,-15.945){0}{\usebox{\plotpoint}}
\put(377.83,296.17){\usebox{\plotpoint}}
\multiput(378,296)(13.287,-15.945){0}{\usebox{\plotpoint}}
\multiput(383,290)(15.945,-13.287){0}{\usebox{\plotpoint}}
\put(392.46,281.54){\usebox{\plotpoint}}
\multiput(394,280)(14.676,-14.676){0}{\usebox{\plotpoint}}
\multiput(399,275)(14.676,-14.676){0}{\usebox{\plotpoint}}
\put(407.69,267.54){\usebox{\plotpoint}}
\multiput(410,266)(14.676,-14.676){0}{\usebox{\plotpoint}}
\multiput(415,261)(16.207,-12.966){0}{\usebox{\plotpoint}}
\put(423.46,254.12){\usebox{\plotpoint}}
\multiput(426,252)(16.207,-12.966){0}{\usebox{\plotpoint}}
\multiput(431,248)(16.207,-12.966){0}{\usebox{\plotpoint}}
\put(439.62,241.10){\usebox{\plotpoint}}
\multiput(441,240)(18.564,-9.282){0}{\usebox{\plotpoint}}
\multiput(447,237)(16.207,-12.966){0}{\usebox{\plotpoint}}
\put(456.59,229.33){\usebox{\plotpoint}}
\multiput(457,229)(17.798,-10.679){0}{\usebox{\plotpoint}}
\multiput(462,226)(18.564,-9.282){0}{\usebox{\plotpoint}}
\multiput(468,223)(16.207,-12.966){0}{\usebox{\plotpoint}}
\put(474.11,218.34){\usebox{\plotpoint}}
\multiput(478,216)(17.798,-10.679){0}{\usebox{\plotpoint}}
\multiput(483,213)(18.564,-9.282){0}{\usebox{\plotpoint}}
\put(492.15,208.11){\usebox{\plotpoint}}
\multiput(494,207)(19.271,-7.708){0}{\usebox{\plotpoint}}
\multiput(499,205)(18.564,-9.282){0}{\usebox{\plotpoint}}
\multiput(505,202)(17.798,-10.679){0}{\usebox{\plotpoint}}
\put(510.63,198.75){\usebox{\plotpoint}}
\multiput(515,197)(17.798,-10.679){0}{\usebox{\plotpoint}}
\multiput(520,194)(19.690,-6.563){0}{\usebox{\plotpoint}}
\put(529.61,190.56){\usebox{\plotpoint}}
\multiput(531,190)(17.798,-10.679){0}{\usebox{\plotpoint}}
\multiput(536,187)(19.271,-7.708){0}{\usebox{\plotpoint}}
\multiput(541,185)(19.690,-6.563){0}{\usebox{\plotpoint}}
\put(548.60,182.36){\usebox{\plotpoint}}
\multiput(552,181)(19.271,-7.708){0}{\usebox{\plotpoint}}
\multiput(557,179)(19.690,-6.563){0}{\usebox{\plotpoint}}
\put(568.00,175.00){\usebox{\plotpoint}}
\multiput(568,175)(19.271,-7.708){0}{\usebox{\plotpoint}}
\multiput(573,173)(19.271,-7.708){0}{\usebox{\plotpoint}}
\multiput(578,171)(20.473,-3.412){0}{\usebox{\plotpoint}}
\put(587.62,168.55){\usebox{\plotpoint}}
\multiput(589,168)(19.271,-7.708){0}{\usebox{\plotpoint}}
\multiput(594,166)(20.352,-4.070){0}{\usebox{\plotpoint}}
\multiput(599,165)(19.690,-6.563){0}{\usebox{\plotpoint}}
\put(607.41,162.52){\usebox{\plotpoint}}
\multiput(610,162)(19.271,-7.708){0}{\usebox{\plotpoint}}
\multiput(615,160)(20.473,-3.412){0}{\usebox{\plotpoint}}
\multiput(621,159)(20.352,-4.070){0}{\usebox{\plotpoint}}
\put(627.44,157.43){\usebox{\plotpoint}}
\multiput(631,156)(20.352,-4.070){0}{\usebox{\plotpoint}}
\multiput(636,155)(20.473,-3.412){0}{\usebox{\plotpoint}}
\multiput(642,154)(20.352,-4.070){0}{\usebox{\plotpoint}}
\put(647.59,152.76){\usebox{\plotpoint}}
\multiput(652,151)(20.352,-4.070){0}{\usebox{\plotpoint}}
\multiput(657,150)(20.473,-3.412){0}{\usebox{\plotpoint}}
\put(667.73,148.05){\usebox{\plotpoint}}
\multiput(668,148)(20.352,-4.070){0}{\usebox{\plotpoint}}
\multiput(673,147)(20.352,-4.070){0}{\usebox{\plotpoint}}
\multiput(678,146)(20.473,-3.412){0}{\usebox{\plotpoint}}
\put(688.12,144.18){\usebox{\plotpoint}}
\multiput(689,144)(20.352,-4.070){0}{\usebox{\plotpoint}}
\multiput(694,143)(20.473,-3.412){0}{\usebox{\plotpoint}}
\multiput(700,142)(20.352,-4.070){0}{\usebox{\plotpoint}}
\put(708.58,141.00){\usebox{\plotpoint}}
\multiput(710,141)(20.352,-4.070){0}{\usebox{\plotpoint}}
\multiput(715,140)(20.473,-3.412){0}{\usebox{\plotpoint}}
\multiput(721,139)(20.352,-4.070){0}{\usebox{\plotpoint}}
\put(728.99,137.40){\usebox{\plotpoint}}
\multiput(731,137)(20.755,0.000){0}{\usebox{\plotpoint}}
\multiput(736,137)(20.473,-3.412){0}{\usebox{\plotpoint}}
\multiput(742,136)(20.352,-4.070){0}{\usebox{\plotpoint}}
\put(747,135){\usebox{\plotpoint}}
\put(220,113){\usebox{\plotpoint}}
\multiput(220.59,113.00)(0.477,2.269){7}{\rule{0.115pt}{1.780pt}}
\multiput(219.17,113.00)(5.000,17.306){2}{\rule{0.400pt}{0.890pt}}
\multiput(225.59,134.00)(0.482,1.847){9}{\rule{0.116pt}{1.500pt}}
\multiput(224.17,134.00)(6.000,17.887){2}{\rule{0.400pt}{0.750pt}}
\multiput(231.59,155.00)(0.477,2.269){7}{\rule{0.115pt}{1.780pt}}
\multiput(230.17,155.00)(5.000,17.306){2}{\rule{0.400pt}{0.890pt}}
\multiput(236.59,176.00)(0.477,2.269){7}{\rule{0.115pt}{1.780pt}}
\multiput(235.17,176.00)(5.000,17.306){2}{\rule{0.400pt}{0.890pt}}
\multiput(241.59,197.00)(0.477,2.157){7}{\rule{0.115pt}{1.700pt}}
\multiput(240.17,197.00)(5.000,16.472){2}{\rule{0.400pt}{0.850pt}}
\multiput(246.59,217.00)(0.482,1.756){9}{\rule{0.116pt}{1.433pt}}
\multiput(245.17,217.00)(6.000,17.025){2}{\rule{0.400pt}{0.717pt}}
\multiput(252.59,237.00)(0.477,2.046){7}{\rule{0.115pt}{1.620pt}}
\multiput(251.17,237.00)(5.000,15.638){2}{\rule{0.400pt}{0.810pt}}
\multiput(257.59,256.00)(0.477,2.046){7}{\rule{0.115pt}{1.620pt}}
\multiput(256.17,256.00)(5.000,15.638){2}{\rule{0.400pt}{0.810pt}}
\multiput(262.59,275.00)(0.477,1.935){7}{\rule{0.115pt}{1.540pt}}
\multiput(261.17,275.00)(5.000,14.804){2}{\rule{0.400pt}{0.770pt}}
\multiput(267.59,293.00)(0.482,1.485){9}{\rule{0.116pt}{1.233pt}}
\multiput(266.17,293.00)(6.000,14.440){2}{\rule{0.400pt}{0.617pt}}
\multiput(273.59,310.00)(0.477,1.823){7}{\rule{0.115pt}{1.460pt}}
\multiput(272.17,310.00)(5.000,13.970){2}{\rule{0.400pt}{0.730pt}}
\multiput(278.59,327.00)(0.477,1.601){7}{\rule{0.115pt}{1.300pt}}
\multiput(277.17,327.00)(5.000,12.302){2}{\rule{0.400pt}{0.650pt}}
\multiput(283.59,342.00)(0.482,1.304){9}{\rule{0.116pt}{1.100pt}}
\multiput(282.17,342.00)(6.000,12.717){2}{\rule{0.400pt}{0.550pt}}
\multiput(289.59,357.00)(0.477,1.489){7}{\rule{0.115pt}{1.220pt}}
\multiput(288.17,357.00)(5.000,11.468){2}{\rule{0.400pt}{0.610pt}}
\multiput(294.59,371.00)(0.477,1.378){7}{\rule{0.115pt}{1.140pt}}
\multiput(293.17,371.00)(5.000,10.634){2}{\rule{0.400pt}{0.570pt}}
\multiput(299.59,384.00)(0.477,1.267){7}{\rule{0.115pt}{1.060pt}}
\multiput(298.17,384.00)(5.000,9.800){2}{\rule{0.400pt}{0.530pt}}
\multiput(304.59,396.00)(0.482,0.852){9}{\rule{0.116pt}{0.767pt}}
\multiput(303.17,396.00)(6.000,8.409){2}{\rule{0.400pt}{0.383pt}}
\multiput(310.59,406.00)(0.477,1.044){7}{\rule{0.115pt}{0.900pt}}
\multiput(309.17,406.00)(5.000,8.132){2}{\rule{0.400pt}{0.450pt}}
\multiput(315.59,416.00)(0.477,0.933){7}{\rule{0.115pt}{0.820pt}}
\multiput(314.17,416.00)(5.000,7.298){2}{\rule{0.400pt}{0.410pt}}
\multiput(320.59,425.00)(0.477,0.710){7}{\rule{0.115pt}{0.660pt}}
\multiput(319.17,425.00)(5.000,5.630){2}{\rule{0.400pt}{0.330pt}}
\multiput(325.00,432.59)(0.491,0.482){9}{\rule{0.500pt}{0.116pt}}
\multiput(325.00,431.17)(4.962,6.000){2}{\rule{0.250pt}{0.400pt}}
\multiput(331.59,438.00)(0.477,0.599){7}{\rule{0.115pt}{0.580pt}}
\multiput(330.17,438.00)(5.000,4.796){2}{\rule{0.400pt}{0.290pt}}
\multiput(336.00,444.60)(0.627,0.468){5}{\rule{0.600pt}{0.113pt}}
\multiput(336.00,443.17)(3.755,4.000){2}{\rule{0.300pt}{0.400pt}}
\multiput(341.00,448.61)(0.909,0.447){3}{\rule{0.767pt}{0.108pt}}
\multiput(341.00,447.17)(3.409,3.000){2}{\rule{0.383pt}{0.400pt}}
\put(346,451.17){\rule{1.300pt}{0.400pt}}
\multiput(346.00,450.17)(3.302,2.000){2}{\rule{0.650pt}{0.400pt}}
\put(352,452.67){\rule{1.204pt}{0.400pt}}
\multiput(352.00,452.17)(2.500,1.000){2}{\rule{0.602pt}{0.400pt}}
\put(368,452.17){\rule{1.100pt}{0.400pt}}
\multiput(368.00,453.17)(2.717,-2.000){2}{\rule{0.550pt}{0.400pt}}
\multiput(373.00,450.95)(0.909,-0.447){3}{\rule{0.767pt}{0.108pt}}
\multiput(373.00,451.17)(3.409,-3.000){2}{\rule{0.383pt}{0.400pt}}
\multiput(378.00,447.95)(0.909,-0.447){3}{\rule{0.767pt}{0.108pt}}
\multiput(378.00,448.17)(3.409,-3.000){2}{\rule{0.383pt}{0.400pt}}
\multiput(383.00,444.93)(0.599,-0.477){7}{\rule{0.580pt}{0.115pt}}
\multiput(383.00,445.17)(4.796,-5.000){2}{\rule{0.290pt}{0.400pt}}
\multiput(389.00,439.93)(0.487,-0.477){7}{\rule{0.500pt}{0.115pt}}
\multiput(389.00,440.17)(3.962,-5.000){2}{\rule{0.250pt}{0.400pt}}
\multiput(394.00,434.93)(0.487,-0.477){7}{\rule{0.500pt}{0.115pt}}
\multiput(394.00,435.17)(3.962,-5.000){2}{\rule{0.250pt}{0.400pt}}
\multiput(399.59,428.59)(0.477,-0.599){7}{\rule{0.115pt}{0.580pt}}
\multiput(398.17,429.80)(5.000,-4.796){2}{\rule{0.400pt}{0.290pt}}
\multiput(404.59,422.65)(0.482,-0.581){9}{\rule{0.116pt}{0.567pt}}
\multiput(403.17,423.82)(6.000,-5.824){2}{\rule{0.400pt}{0.283pt}}
\multiput(410.59,415.26)(0.477,-0.710){7}{\rule{0.115pt}{0.660pt}}
\multiput(409.17,416.63)(5.000,-5.630){2}{\rule{0.400pt}{0.330pt}}
\multiput(415.59,407.93)(0.477,-0.821){7}{\rule{0.115pt}{0.740pt}}
\multiput(414.17,409.46)(5.000,-6.464){2}{\rule{0.400pt}{0.370pt}}
\multiput(420.59,400.37)(0.482,-0.671){9}{\rule{0.116pt}{0.633pt}}
\multiput(419.17,401.69)(6.000,-6.685){2}{\rule{0.400pt}{0.317pt}}
\multiput(426.59,391.60)(0.477,-0.933){7}{\rule{0.115pt}{0.820pt}}
\multiput(425.17,393.30)(5.000,-7.298){2}{\rule{0.400pt}{0.410pt}}
\multiput(431.59,382.60)(0.477,-0.933){7}{\rule{0.115pt}{0.820pt}}
\multiput(430.17,384.30)(5.000,-7.298){2}{\rule{0.400pt}{0.410pt}}
\multiput(436.59,373.60)(0.477,-0.933){7}{\rule{0.115pt}{0.820pt}}
\multiput(435.17,375.30)(5.000,-7.298){2}{\rule{0.400pt}{0.410pt}}
\multiput(441.59,365.09)(0.482,-0.762){9}{\rule{0.116pt}{0.700pt}}
\multiput(440.17,366.55)(6.000,-7.547){2}{\rule{0.400pt}{0.350pt}}
\multiput(447.59,355.60)(0.477,-0.933){7}{\rule{0.115pt}{0.820pt}}
\multiput(446.17,357.30)(5.000,-7.298){2}{\rule{0.400pt}{0.410pt}}
\multiput(452.59,346.60)(0.477,-0.933){7}{\rule{0.115pt}{0.820pt}}
\multiput(451.17,348.30)(5.000,-7.298){2}{\rule{0.400pt}{0.410pt}}
\multiput(457.59,337.26)(0.477,-1.044){7}{\rule{0.115pt}{0.900pt}}
\multiput(456.17,339.13)(5.000,-8.132){2}{\rule{0.400pt}{0.450pt}}
\multiput(462.59,328.09)(0.482,-0.762){9}{\rule{0.116pt}{0.700pt}}
\multiput(461.17,329.55)(6.000,-7.547){2}{\rule{0.400pt}{0.350pt}}
\multiput(468.59,318.26)(0.477,-1.044){7}{\rule{0.115pt}{0.900pt}}
\multiput(467.17,320.13)(5.000,-8.132){2}{\rule{0.400pt}{0.450pt}}
\multiput(473.59,308.60)(0.477,-0.933){7}{\rule{0.115pt}{0.820pt}}
\multiput(472.17,310.30)(5.000,-7.298){2}{\rule{0.400pt}{0.410pt}}
\multiput(478.59,299.60)(0.477,-0.933){7}{\rule{0.115pt}{0.820pt}}
\multiput(477.17,301.30)(5.000,-7.298){2}{\rule{0.400pt}{0.410pt}}
\multiput(483.59,291.09)(0.482,-0.762){9}{\rule{0.116pt}{0.700pt}}
\multiput(482.17,292.55)(6.000,-7.547){2}{\rule{0.400pt}{0.350pt}}
\multiput(489.59,281.60)(0.477,-0.933){7}{\rule{0.115pt}{0.820pt}}
\multiput(488.17,283.30)(5.000,-7.298){2}{\rule{0.400pt}{0.410pt}}
\multiput(494.59,272.60)(0.477,-0.933){7}{\rule{0.115pt}{0.820pt}}
\multiput(493.17,274.30)(5.000,-7.298){2}{\rule{0.400pt}{0.410pt}}
\multiput(499.59,264.37)(0.482,-0.671){9}{\rule{0.116pt}{0.633pt}}
\multiput(498.17,265.69)(6.000,-6.685){2}{\rule{0.400pt}{0.317pt}}
\multiput(505.59,255.60)(0.477,-0.933){7}{\rule{0.115pt}{0.820pt}}
\multiput(504.17,257.30)(5.000,-7.298){2}{\rule{0.400pt}{0.410pt}}
\multiput(510.59,246.93)(0.477,-0.821){7}{\rule{0.115pt}{0.740pt}}
\multiput(509.17,248.46)(5.000,-6.464){2}{\rule{0.400pt}{0.370pt}}
\multiput(515.59,238.93)(0.477,-0.821){7}{\rule{0.115pt}{0.740pt}}
\multiput(514.17,240.46)(5.000,-6.464){2}{\rule{0.400pt}{0.370pt}}
\multiput(520.59,231.65)(0.482,-0.581){9}{\rule{0.116pt}{0.567pt}}
\multiput(519.17,232.82)(6.000,-5.824){2}{\rule{0.400pt}{0.283pt}}
\multiput(526.59,224.26)(0.477,-0.710){7}{\rule{0.115pt}{0.660pt}}
\multiput(525.17,225.63)(5.000,-5.630){2}{\rule{0.400pt}{0.330pt}}
\multiput(531.59,217.26)(0.477,-0.710){7}{\rule{0.115pt}{0.660pt}}
\multiput(530.17,218.63)(5.000,-5.630){2}{\rule{0.400pt}{0.330pt}}
\multiput(536.59,210.26)(0.477,-0.710){7}{\rule{0.115pt}{0.660pt}}
\multiput(535.17,211.63)(5.000,-5.630){2}{\rule{0.400pt}{0.330pt}}
\multiput(541.00,204.93)(0.491,-0.482){9}{\rule{0.500pt}{0.116pt}}
\multiput(541.00,205.17)(4.962,-6.000){2}{\rule{0.250pt}{0.400pt}}
\multiput(547.59,197.59)(0.477,-0.599){7}{\rule{0.115pt}{0.580pt}}
\multiput(546.17,198.80)(5.000,-4.796){2}{\rule{0.400pt}{0.290pt}}
\multiput(552.59,191.59)(0.477,-0.599){7}{\rule{0.115pt}{0.580pt}}
\multiput(551.17,192.80)(5.000,-4.796){2}{\rule{0.400pt}{0.290pt}}
\multiput(557.00,186.93)(0.491,-0.482){9}{\rule{0.500pt}{0.116pt}}
\multiput(557.00,187.17)(4.962,-6.000){2}{\rule{0.250pt}{0.400pt}}
\multiput(563.00,180.93)(0.487,-0.477){7}{\rule{0.500pt}{0.115pt}}
\multiput(563.00,181.17)(3.962,-5.000){2}{\rule{0.250pt}{0.400pt}}
\multiput(568.00,175.93)(0.487,-0.477){7}{\rule{0.500pt}{0.115pt}}
\multiput(568.00,176.17)(3.962,-5.000){2}{\rule{0.250pt}{0.400pt}}
\multiput(573.00,170.94)(0.627,-0.468){5}{\rule{0.600pt}{0.113pt}}
\multiput(573.00,171.17)(3.755,-4.000){2}{\rule{0.300pt}{0.400pt}}
\multiput(578.00,166.93)(0.599,-0.477){7}{\rule{0.580pt}{0.115pt}}
\multiput(578.00,167.17)(4.796,-5.000){2}{\rule{0.290pt}{0.400pt}}
\multiput(584.00,161.94)(0.627,-0.468){5}{\rule{0.600pt}{0.113pt}}
\multiput(584.00,162.17)(3.755,-4.000){2}{\rule{0.300pt}{0.400pt}}
\multiput(589.00,157.95)(0.909,-0.447){3}{\rule{0.767pt}{0.108pt}}
\multiput(589.00,158.17)(3.409,-3.000){2}{\rule{0.383pt}{0.400pt}}
\multiput(594.00,154.94)(0.627,-0.468){5}{\rule{0.600pt}{0.113pt}}
\multiput(594.00,155.17)(3.755,-4.000){2}{\rule{0.300pt}{0.400pt}}
\multiput(599.00,150.95)(1.132,-0.447){3}{\rule{0.900pt}{0.108pt}}
\multiput(599.00,151.17)(4.132,-3.000){2}{\rule{0.450pt}{0.400pt}}
\multiput(605.00,147.95)(0.909,-0.447){3}{\rule{0.767pt}{0.108pt}}
\multiput(605.00,148.17)(3.409,-3.000){2}{\rule{0.383pt}{0.400pt}}
\multiput(610.00,144.95)(0.909,-0.447){3}{\rule{0.767pt}{0.108pt}}
\multiput(610.00,145.17)(3.409,-3.000){2}{\rule{0.383pt}{0.400pt}}
\multiput(615.00,141.95)(1.132,-0.447){3}{\rule{0.900pt}{0.108pt}}
\multiput(615.00,142.17)(4.132,-3.000){2}{\rule{0.450pt}{0.400pt}}
\put(621,138.17){\rule{1.100pt}{0.400pt}}
\multiput(621.00,139.17)(2.717,-2.000){2}{\rule{0.550pt}{0.400pt}}
\multiput(626.00,136.95)(0.909,-0.447){3}{\rule{0.767pt}{0.108pt}}
\multiput(626.00,137.17)(3.409,-3.000){2}{\rule{0.383pt}{0.400pt}}
\put(631,133.17){\rule{1.100pt}{0.400pt}}
\multiput(631.00,134.17)(2.717,-2.000){2}{\rule{0.550pt}{0.400pt}}
\put(636,131.17){\rule{1.300pt}{0.400pt}}
\multiput(636.00,132.17)(3.302,-2.000){2}{\rule{0.650pt}{0.400pt}}
\put(642,129.67){\rule{1.204pt}{0.400pt}}
\multiput(642.00,130.17)(2.500,-1.000){2}{\rule{0.602pt}{0.400pt}}
\put(647,128.17){\rule{1.100pt}{0.400pt}}
\multiput(647.00,129.17)(2.717,-2.000){2}{\rule{0.550pt}{0.400pt}}
\put(652,126.17){\rule{1.100pt}{0.400pt}}
\multiput(652.00,127.17)(2.717,-2.000){2}{\rule{0.550pt}{0.400pt}}
\put(657,124.67){\rule{1.445pt}{0.400pt}}
\multiput(657.00,125.17)(3.000,-1.000){2}{\rule{0.723pt}{0.400pt}}
\put(663,123.67){\rule{1.204pt}{0.400pt}}
\multiput(663.00,124.17)(2.500,-1.000){2}{\rule{0.602pt}{0.400pt}}
\put(668,122.67){\rule{1.204pt}{0.400pt}}
\multiput(668.00,123.17)(2.500,-1.000){2}{\rule{0.602pt}{0.400pt}}
\put(673,121.67){\rule{1.204pt}{0.400pt}}
\multiput(673.00,122.17)(2.500,-1.000){2}{\rule{0.602pt}{0.400pt}}
\put(678,120.67){\rule{1.445pt}{0.400pt}}
\multiput(678.00,121.17)(3.000,-1.000){2}{\rule{0.723pt}{0.400pt}}
\put(684,119.67){\rule{1.204pt}{0.400pt}}
\multiput(684.00,120.17)(2.500,-1.000){2}{\rule{0.602pt}{0.400pt}}
\put(689,118.67){\rule{1.204pt}{0.400pt}}
\multiput(689.00,119.17)(2.500,-1.000){2}{\rule{0.602pt}{0.400pt}}
\put(357.0,454.0){\rule[-0.200pt]{2.650pt}{0.400pt}}
\put(700,117.67){\rule{1.204pt}{0.400pt}}
\multiput(700.00,118.17)(2.500,-1.000){2}{\rule{0.602pt}{0.400pt}}
\put(705,116.67){\rule{1.204pt}{0.400pt}}
\multiput(705.00,117.17)(2.500,-1.000){2}{\rule{0.602pt}{0.400pt}}
\put(694.0,119.0){\rule[-0.200pt]{1.445pt}{0.400pt}}
\put(715,115.67){\rule{1.445pt}{0.400pt}}
\multiput(715.00,116.17)(3.000,-1.000){2}{\rule{0.723pt}{0.400pt}}
\put(710.0,117.0){\rule[-0.200pt]{1.204pt}{0.400pt}}
\put(731,114.67){\rule{1.204pt}{0.400pt}}
\multiput(731.00,115.17)(2.500,-1.000){2}{\rule{0.602pt}{0.400pt}}
\put(721.0,116.0){\rule[-0.200pt]{2.409pt}{0.400pt}}
\put(736.0,115.0){\rule[-0.200pt]{2.650pt}{0.400pt}}
\put(659,562){\makebox(0,0)[r]{$v_{\theta}^{-1}=10$}}
\put(681.0,562.0){\rule[-0.200pt]{15.899pt}{0.400pt}}
\put(220.0,113.0){\rule[-0.200pt]{0.400pt}{16.140pt}}
\put(220.0,180.0){\rule[-0.200pt]{4.336pt}{0.400pt}}
\put(238.0,113.0){\rule[-0.200pt]{0.400pt}{16.140pt}}
\put(220.0,113.0){\rule[-0.200pt]{4.336pt}{0.400pt}}
\put(238.0,113.0){\rule[-0.200pt]{0.400pt}{44.085pt}}
\put(238.0,296.0){\rule[-0.200pt]{4.095pt}{0.400pt}}
\put(255.0,113.0){\rule[-0.200pt]{0.400pt}{44.085pt}}
\put(238.0,113.0){\rule[-0.200pt]{4.095pt}{0.400pt}}
\put(255.0,113.0){\rule[-0.200pt]{0.400pt}{78.533pt}}
\put(255.0,439.0){\rule[-0.200pt]{4.336pt}{0.400pt}}
\put(273.0,113.0){\rule[-0.200pt]{0.400pt}{78.533pt}}
\put(255.0,113.0){\rule[-0.200pt]{4.336pt}{0.400pt}}
\put(273.0,113.0){\rule[-0.200pt]{0.400pt}{111.778pt}}
\put(273.0,577.0){\rule[-0.200pt]{4.095pt}{0.400pt}}
\put(290.0,113.0){\rule[-0.200pt]{0.400pt}{111.778pt}}
\put(273.0,113.0){\rule[-0.200pt]{4.095pt}{0.400pt}}
\put(290.0,113.0){\rule[-0.200pt]{0.400pt}{95.878pt}}
\put(290.0,511.0){\rule[-0.200pt]{4.336pt}{0.400pt}}
\put(308.0,113.0){\rule[-0.200pt]{0.400pt}{95.878pt}}
\put(290.0,113.0){\rule[-0.200pt]{4.336pt}{0.400pt}}
\put(308.0,113.0){\rule[-0.200pt]{0.400pt}{85.038pt}}
\put(308.0,466.0){\rule[-0.200pt]{4.095pt}{0.400pt}}
\put(325.0,113.0){\rule[-0.200pt]{0.400pt}{85.038pt}}
\put(308.0,113.0){\rule[-0.200pt]{4.095pt}{0.400pt}}
\put(325.0,113.0){\rule[-0.200pt]{0.400pt}{85.519pt}}
\put(325.0,468.0){\rule[-0.200pt]{4.336pt}{0.400pt}}
\put(343.0,113.0){\rule[-0.200pt]{0.400pt}{85.519pt}}
\put(325.0,113.0){\rule[-0.200pt]{4.336pt}{0.400pt}}
\put(343.0,113.0){\rule[-0.200pt]{0.400pt}{70.343pt}}
\put(343.0,405.0){\rule[-0.200pt]{4.336pt}{0.400pt}}
\put(361.0,113.0){\rule[-0.200pt]{0.400pt}{70.343pt}}
\put(343.0,113.0){\rule[-0.200pt]{4.336pt}{0.400pt}}
\put(361.0,113.0){\rule[-0.200pt]{0.400pt}{69.138pt}}
\put(361.0,400.0){\rule[-0.200pt]{4.095pt}{0.400pt}}
\put(378.0,113.0){\rule[-0.200pt]{0.400pt}{69.138pt}}
\put(361.0,113.0){\rule[-0.200pt]{4.095pt}{0.400pt}}
\put(378.0,113.0){\rule[-0.200pt]{0.400pt}{50.107pt}}
\put(378.0,321.0){\rule[-0.200pt]{4.336pt}{0.400pt}}
\put(396.0,113.0){\rule[-0.200pt]{0.400pt}{50.107pt}}
\put(378.0,113.0){\rule[-0.200pt]{4.336pt}{0.400pt}}
\put(396.0,113.0){\rule[-0.200pt]{0.400pt}{37.821pt}}
\put(396.0,270.0){\rule[-0.200pt]{4.095pt}{0.400pt}}
\put(413.0,113.0){\rule[-0.200pt]{0.400pt}{37.821pt}}
\put(396.0,113.0){\rule[-0.200pt]{4.095pt}{0.400pt}}
\put(413.0,113.0){\rule[-0.200pt]{0.400pt}{34.930pt}}
\put(413.0,258.0){\rule[-0.200pt]{4.336pt}{0.400pt}}
\put(431.0,113.0){\rule[-0.200pt]{0.400pt}{34.930pt}}
\put(413.0,113.0){\rule[-0.200pt]{4.336pt}{0.400pt}}
\put(431.0,113.0){\rule[-0.200pt]{0.400pt}{33.967pt}}
\put(431.0,254.0){\rule[-0.200pt]{4.095pt}{0.400pt}}
\put(448.0,113.0){\rule[-0.200pt]{0.400pt}{33.967pt}}
\put(431.0,113.0){\rule[-0.200pt]{4.095pt}{0.400pt}}
\put(448.0,113.0){\rule[-0.200pt]{0.400pt}{23.849pt}}
\put(448.0,212.0){\rule[-0.200pt]{4.336pt}{0.400pt}}
\put(466.0,113.0){\rule[-0.200pt]{0.400pt}{23.849pt}}
\put(448.0,113.0){\rule[-0.200pt]{4.336pt}{0.400pt}}
\put(466.0,113.0){\rule[-0.200pt]{0.400pt}{31.799pt}}
\put(466.0,245.0){\rule[-0.200pt]{4.095pt}{0.400pt}}
\put(483.0,113.0){\rule[-0.200pt]{0.400pt}{31.799pt}}
\put(466.0,113.0){\rule[-0.200pt]{4.095pt}{0.400pt}}
\put(483.0,113.0){\rule[-0.200pt]{0.400pt}{22.404pt}}
\put(483.0,206.0){\rule[-0.200pt]{4.336pt}{0.400pt}}
\put(501.0,113.0){\rule[-0.200pt]{0.400pt}{22.404pt}}
\put(483.0,113.0){\rule[-0.200pt]{4.336pt}{0.400pt}}
\put(501.0,113.0){\rule[-0.200pt]{0.400pt}{22.404pt}}
\put(501.0,206.0){\rule[-0.200pt]{4.336pt}{0.400pt}}
\put(519.0,113.0){\rule[-0.200pt]{0.400pt}{22.404pt}}
\put(501.0,113.0){\rule[-0.200pt]{4.336pt}{0.400pt}}
\put(519.0,113.0){\rule[-0.200pt]{0.400pt}{13.731pt}}
\put(519.0,170.0){\rule[-0.200pt]{4.095pt}{0.400pt}}
\put(536.0,113.0){\rule[-0.200pt]{0.400pt}{13.731pt}}
\put(519.0,113.0){\rule[-0.200pt]{4.095pt}{0.400pt}}
\put(536.0,113.0){\rule[-0.200pt]{0.400pt}{18.308pt}}
\put(536.0,189.0){\rule[-0.200pt]{4.336pt}{0.400pt}}
\put(554.0,113.0){\rule[-0.200pt]{0.400pt}{18.308pt}}
\put(536.0,113.0){\rule[-0.200pt]{4.336pt}{0.400pt}}
\put(554.0,113.0){\rule[-0.200pt]{0.400pt}{15.899pt}}
\put(554.0,179.0){\rule[-0.200pt]{4.095pt}{0.400pt}}
\put(571.0,113.0){\rule[-0.200pt]{0.400pt}{15.899pt}}
\put(554.0,113.0){\rule[-0.200pt]{4.095pt}{0.400pt}}
\put(571.0,113.0){\rule[-0.200pt]{0.400pt}{11.563pt}}
\put(571.0,161.0){\rule[-0.200pt]{4.336pt}{0.400pt}}
\put(589.0,113.0){\rule[-0.200pt]{0.400pt}{11.563pt}}
\put(571.0,113.0){\rule[-0.200pt]{4.336pt}{0.400pt}}
\put(589.0,113.0){\rule[-0.200pt]{0.400pt}{8.913pt}}
\put(589.0,150.0){\rule[-0.200pt]{4.095pt}{0.400pt}}
\put(606.0,113.0){\rule[-0.200pt]{0.400pt}{8.913pt}}
\put(589.0,113.0){\rule[-0.200pt]{4.095pt}{0.400pt}}
\put(606.0,113.0){\rule[-0.200pt]{0.400pt}{13.972pt}}
\put(606.0,171.0){\rule[-0.200pt]{4.336pt}{0.400pt}}
\put(624.0,113.0){\rule[-0.200pt]{0.400pt}{13.972pt}}
\put(606.0,113.0){\rule[-0.200pt]{4.336pt}{0.400pt}}
\put(624.0,113.0){\rule[-0.200pt]{0.400pt}{8.913pt}}
\put(624.0,150.0){\rule[-0.200pt]{4.336pt}{0.400pt}}
\put(642.0,113.0){\rule[-0.200pt]{0.400pt}{8.913pt}}
\put(624.0,113.0){\rule[-0.200pt]{4.336pt}{0.400pt}}
\put(642.0,113.0){\rule[-0.200pt]{0.400pt}{6.022pt}}
\put(642.0,138.0){\rule[-0.200pt]{4.095pt}{0.400pt}}
\put(659.0,113.0){\rule[-0.200pt]{0.400pt}{6.022pt}}
\put(642.0,113.0){\rule[-0.200pt]{4.095pt}{0.400pt}}
\put(659.0,113.0){\rule[-0.200pt]{0.400pt}{7.227pt}}
\put(659.0,143.0){\rule[-0.200pt]{4.336pt}{0.400pt}}
\put(677.0,113.0){\rule[-0.200pt]{0.400pt}{7.227pt}}
\put(659.0,113.0){\rule[-0.200pt]{4.336pt}{0.400pt}}
\put(677.0,113.0){\rule[-0.200pt]{0.400pt}{5.059pt}}
\put(677.0,134.0){\rule[-0.200pt]{4.095pt}{0.400pt}}
\put(694.0,113.0){\rule[-0.200pt]{0.400pt}{5.059pt}}
\put(677.0,113.0){\rule[-0.200pt]{4.095pt}{0.400pt}}
\put(694.0,113.0){\rule[-0.200pt]{0.400pt}{4.336pt}}
\put(694.0,131.0){\rule[-0.200pt]{4.336pt}{0.400pt}}
\put(712.0,113.0){\rule[-0.200pt]{0.400pt}{4.336pt}}
\put(694.0,113.0){\rule[-0.200pt]{4.336pt}{0.400pt}}
\put(712.0,113.0){\rule[-0.200pt]{0.400pt}{4.577pt}}
\put(712.0,132.0){\rule[-0.200pt]{4.095pt}{0.400pt}}
\put(729.0,113.0){\rule[-0.200pt]{0.400pt}{4.577pt}}
\put(712.0,113.0){\rule[-0.200pt]{4.095pt}{0.400pt}}
\put(729.0,113.0){\rule[-0.200pt]{0.400pt}{5.300pt}}
\put(729.0,135.0){\rule[-0.200pt]{4.336pt}{0.400pt}}
\put(747.0,113.0){\rule[-0.200pt]{0.400pt}{5.300pt}}
\put(729.0,113.0){\rule[-0.200pt]{4.336pt}{0.400pt}}
\end{picture}

%% file: ps_lv30.tex
\setlength{\unitlength}{0.240900pt}
\ifx\plotpoint\undefined\newsavebox{\plotpoint}\fi
\sbox{\plotpoint}{\rule[-0.200pt]{0.400pt}{0.400pt}}%
\begin{picture}(811,675)(0,0)
\font\gnuplot=cmr10 at 10pt
\gnuplot
\input{ps_inc}
\put(220.0,113.0){\rule[-0.200pt]{126.954pt}{0.400pt}}
\put(220.0,113.0){\rule[-0.200pt]{0.400pt}{129.845pt}}
\put(220.0,113.0){\rule[-0.200pt]{4.818pt}{0.400pt}}
\put(198,113){\makebox(0,0)[r]{0}}
\put(727.0,113.0){\rule[-0.200pt]{4.818pt}{0.400pt}}
\put(220.0,203.0){\rule[-0.200pt]{4.818pt}{0.400pt}}
\put(198,203){\makebox(0,0)[r]{0.2}}
\put(727.0,203.0){\rule[-0.200pt]{4.818pt}{0.400pt}}
\put(220.0,293.0){\rule[-0.200pt]{4.818pt}{0.400pt}}
\put(198,293){\makebox(0,0)[r]{0.4}}
\put(727.0,293.0){\rule[-0.200pt]{4.818pt}{0.400pt}}
\put(220.0,382.0){\rule[-0.200pt]{4.818pt}{0.400pt}}
\put(198,382){\makebox(0,0)[r]{0.6}}
\put(727.0,382.0){\rule[-0.200pt]{4.818pt}{0.400pt}}
\put(220.0,472.0){\rule[-0.200pt]{4.818pt}{0.400pt}}
\put(198,472){\makebox(0,0)[r]{0.8}}
\put(727.0,472.0){\rule[-0.200pt]{4.818pt}{0.400pt}}
\put(220.0,562.0){\rule[-0.200pt]{4.818pt}{0.400pt}}
\put(198,562){\makebox(0,0)[r]{1}}
\put(727.0,562.0){\rule[-0.200pt]{4.818pt}{0.400pt}}
\put(220.0,652.0){\rule[-0.200pt]{4.818pt}{0.400pt}}
\put(198,652){\makebox(0,0)[r]{1.2}}
\put(727.0,652.0){\rule[-0.200pt]{4.818pt}{0.400pt}}
\put(220.0,113.0){\rule[-0.200pt]{0.400pt}{4.818pt}}
\put(220,68){\makebox(0,0){0}}
\put(220.0,632.0){\rule[-0.200pt]{0.400pt}{4.818pt}}
\put(308.0,113.0){\rule[-0.200pt]{0.400pt}{4.818pt}}
\put(308,68){\makebox(0,0){0.5}}
\put(308.0,632.0){\rule[-0.200pt]{0.400pt}{4.818pt}}
\put(396.0,113.0){\rule[-0.200pt]{0.400pt}{4.818pt}}
\put(396,68){\makebox(0,0){1}}
\put(396.0,632.0){\rule[-0.200pt]{0.400pt}{4.818pt}}
\put(483.0,113.0){\rule[-0.200pt]{0.400pt}{4.818pt}}
\put(483,68){\makebox(0,0){1.5}}
\put(483.0,632.0){\rule[-0.200pt]{0.400pt}{4.818pt}}
\put(571.0,113.0){\rule[-0.200pt]{0.400pt}{4.818pt}}
\put(571,68){\makebox(0,0){2}}
\put(571.0,632.0){\rule[-0.200pt]{0.400pt}{4.818pt}}
\put(659.0,113.0){\rule[-0.200pt]{0.400pt}{4.818pt}}
\put(659,68){\makebox(0,0){2.5}}
\put(659.0,632.0){\rule[-0.200pt]{0.400pt}{4.818pt}}
\put(747.0,113.0){\rule[-0.200pt]{0.400pt}{4.818pt}}
\put(747,68){\makebox(0,0){3}}
\put(747.0,632.0){\rule[-0.200pt]{0.400pt}{4.818pt}}
\put(220.0,113.0){\rule[-0.200pt]{126.954pt}{0.400pt}}
\put(747.0,113.0){\rule[-0.200pt]{0.400pt}{129.845pt}}
\put(220.0,652.0){\rule[-0.200pt]{126.954pt}{0.400pt}}
\put(45,382){\makebox(0,0){$P(S)$}}
\put(483,23){\makebox(0,0){$S$}}
\put(220.0,113.0){\rule[-0.200pt]{0.400pt}{129.845pt}}
\put(220,562){\usebox{\plotpoint}}
\put(220.00,562.00){\usebox{\plotpoint}}
\put(227.86,542.80){\usebox{\plotpoint}}
\multiput(231,536)(7.983,-19.159){0}{\usebox{\plotpoint}}
\put(236.10,523.75){\usebox{\plotpoint}}
\put(243.93,504.54){\usebox{\plotpoint}}
\multiput(246,500)(9.282,-18.564){0}{\usebox{\plotpoint}}
\put(252.97,485.86){\usebox{\plotpoint}}
\put(261.56,466.97){\usebox{\plotpoint}}
\multiput(262,466)(9.282,-18.564){0}{\usebox{\plotpoint}}
\put(271.38,448.70){\usebox{\plotpoint}}
\multiput(273,446)(9.282,-18.564){0}{\usebox{\plotpoint}}
\put(280.87,430.25){\usebox{\plotpoint}}
\multiput(283,426)(11.513,-17.270){0}{\usebox{\plotpoint}}
\put(291.52,412.47){\usebox{\plotpoint}}
\multiput(294,408)(10.080,-18.144){0}{\usebox{\plotpoint}}
\put(301.83,394.47){\usebox{\plotpoint}}
\multiput(304,391)(12.453,-16.604){0}{\usebox{\plotpoint}}
\put(313.53,377.35){\usebox{\plotpoint}}
\multiput(315,375)(11.000,-17.601){0}{\usebox{\plotpoint}}
\put(324.97,360.04){\usebox{\plotpoint}}
\multiput(325,360)(12.453,-16.604){0}{\usebox{\plotpoint}}
\multiput(331,352)(12.064,-16.889){0}{\usebox{\plotpoint}}
\put(337.22,343.29){\usebox{\plotpoint}}
\multiput(341,338)(13.287,-15.945){0}{\usebox{\plotpoint}}
\put(350.20,327.10){\usebox{\plotpoint}}
\multiput(352,325)(13.287,-15.945){0}{\usebox{\plotpoint}}
\multiput(357,319)(13.287,-15.945){0}{\usebox{\plotpoint}}
\put(363.67,311.33){\usebox{\plotpoint}}
\multiput(368,307)(13.287,-15.945){0}{\usebox{\plotpoint}}
\put(377.83,296.17){\usebox{\plotpoint}}
\multiput(378,296)(13.287,-15.945){0}{\usebox{\plotpoint}}
\multiput(383,290)(15.945,-13.287){0}{\usebox{\plotpoint}}
\put(392.46,281.54){\usebox{\plotpoint}}
\multiput(394,280)(14.676,-14.676){0}{\usebox{\plotpoint}}
\multiput(399,275)(14.676,-14.676){0}{\usebox{\plotpoint}}
\put(407.69,267.54){\usebox{\plotpoint}}
\multiput(410,266)(14.676,-14.676){0}{\usebox{\plotpoint}}
\multiput(415,261)(16.207,-12.966){0}{\usebox{\plotpoint}}
\put(423.46,254.12){\usebox{\plotpoint}}
\multiput(426,252)(16.207,-12.966){0}{\usebox{\plotpoint}}
\multiput(431,248)(16.207,-12.966){0}{\usebox{\plotpoint}}
\put(439.62,241.10){\usebox{\plotpoint}}
\multiput(441,240)(18.564,-9.282){0}{\usebox{\plotpoint}}
\multiput(447,237)(16.207,-12.966){0}{\usebox{\plotpoint}}
\put(456.59,229.33){\usebox{\plotpoint}}
\multiput(457,229)(17.798,-10.679){0}{\usebox{\plotpoint}}
\multiput(462,226)(18.564,-9.282){0}{\usebox{\plotpoint}}
\multiput(468,223)(16.207,-12.966){0}{\usebox{\plotpoint}}
\put(474.11,218.34){\usebox{\plotpoint}}
\multiput(478,216)(17.798,-10.679){0}{\usebox{\plotpoint}}
\multiput(483,213)(18.564,-9.282){0}{\usebox{\plotpoint}}
\put(492.15,208.11){\usebox{\plotpoint}}
\multiput(494,207)(19.271,-7.708){0}{\usebox{\plotpoint}}
\multiput(499,205)(18.564,-9.282){0}{\usebox{\plotpoint}}
\multiput(505,202)(17.798,-10.679){0}{\usebox{\plotpoint}}
\put(510.63,198.75){\usebox{\plotpoint}}
\multiput(515,197)(17.798,-10.679){0}{\usebox{\plotpoint}}
\multiput(520,194)(19.690,-6.563){0}{\usebox{\plotpoint}}
\put(529.61,190.56){\usebox{\plotpoint}}
\multiput(531,190)(17.798,-10.679){0}{\usebox{\plotpoint}}
\multiput(536,187)(19.271,-7.708){0}{\usebox{\plotpoint}}
\multiput(541,185)(19.690,-6.563){0}{\usebox{\plotpoint}}
\put(548.60,182.36){\usebox{\plotpoint}}
\multiput(552,181)(19.271,-7.708){0}{\usebox{\plotpoint}}
\multiput(557,179)(19.690,-6.563){0}{\usebox{\plotpoint}}
\put(568.00,175.00){\usebox{\plotpoint}}
\multiput(568,175)(19.271,-7.708){0}{\usebox{\plotpoint}}
\multiput(573,173)(19.271,-7.708){0}{\usebox{\plotpoint}}
\multiput(578,171)(20.473,-3.412){0}{\usebox{\plotpoint}}
\put(587.62,168.55){\usebox{\plotpoint}}
\multiput(589,168)(19.271,-7.708){0}{\usebox{\plotpoint}}
\multiput(594,166)(20.352,-4.070){0}{\usebox{\plotpoint}}
\multiput(599,165)(19.690,-6.563){0}{\usebox{\plotpoint}}
\put(607.41,162.52){\usebox{\plotpoint}}
\multiput(610,162)(19.271,-7.708){0}{\usebox{\plotpoint}}
\multiput(615,160)(20.473,-3.412){0}{\usebox{\plotpoint}}
\multiput(621,159)(20.352,-4.070){0}{\usebox{\plotpoint}}
\put(627.44,157.43){\usebox{\plotpoint}}
\multiput(631,156)(20.352,-4.070){0}{\usebox{\plotpoint}}
\multiput(636,155)(20.473,-3.412){0}{\usebox{\plotpoint}}
\multiput(642,154)(20.352,-4.070){0}{\usebox{\plotpoint}}
\put(647.59,152.76){\usebox{\plotpoint}}
\multiput(652,151)(20.352,-4.070){0}{\usebox{\plotpoint}}
\multiput(657,150)(20.473,-3.412){0}{\usebox{\plotpoint}}
\put(667.73,148.05){\usebox{\plotpoint}}
\multiput(668,148)(20.352,-4.070){0}{\usebox{\plotpoint}}
\multiput(673,147)(20.352,-4.070){0}{\usebox{\plotpoint}}
\multiput(678,146)(20.473,-3.412){0}{\usebox{\plotpoint}}
\put(688.12,144.18){\usebox{\plotpoint}}
\multiput(689,144)(20.352,-4.070){0}{\usebox{\plotpoint}}
\multiput(694,143)(20.473,-3.412){0}{\usebox{\plotpoint}}
\multiput(700,142)(20.352,-4.070){0}{\usebox{\plotpoint}}
\put(708.58,141.00){\usebox{\plotpoint}}
\multiput(710,141)(20.352,-4.070){0}{\usebox{\plotpoint}}
\multiput(715,140)(20.473,-3.412){0}{\usebox{\plotpoint}}
\multiput(721,139)(20.352,-4.070){0}{\usebox{\plotpoint}}
\put(728.99,137.40){\usebox{\plotpoint}}
\multiput(731,137)(20.755,0.000){0}{\usebox{\plotpoint}}
\multiput(736,137)(20.473,-3.412){0}{\usebox{\plotpoint}}
\multiput(742,136)(20.352,-4.070){0}{\usebox{\plotpoint}}
\put(747,135){\usebox{\plotpoint}}
\put(220,113){\usebox{\plotpoint}}
\multiput(220.59,113.00)(0.477,2.269){7}{\rule{0.115pt}{1.780pt}}
\multiput(219.17,113.00)(5.000,17.306){2}{\rule{0.400pt}{0.890pt}}
\multiput(225.59,134.00)(0.482,1.847){9}{\rule{0.116pt}{1.500pt}}
\multiput(224.17,134.00)(6.000,17.887){2}{\rule{0.400pt}{0.750pt}}
\multiput(231.59,155.00)(0.477,2.269){7}{\rule{0.115pt}{1.780pt}}
\multiput(230.17,155.00)(5.000,17.306){2}{\rule{0.400pt}{0.890pt}}
\multiput(236.59,176.00)(0.477,2.269){7}{\rule{0.115pt}{1.780pt}}
\multiput(235.17,176.00)(5.000,17.306){2}{\rule{0.400pt}{0.890pt}}
\multiput(241.59,197.00)(0.477,2.157){7}{\rule{0.115pt}{1.700pt}}
\multiput(240.17,197.00)(5.000,16.472){2}{\rule{0.400pt}{0.850pt}}
\multiput(246.59,217.00)(0.482,1.756){9}{\rule{0.116pt}{1.433pt}}
\multiput(245.17,217.00)(6.000,17.025){2}{\rule{0.400pt}{0.717pt}}
\multiput(252.59,237.00)(0.477,2.046){7}{\rule{0.115pt}{1.620pt}}
\multiput(251.17,237.00)(5.000,15.638){2}{\rule{0.400pt}{0.810pt}}
\multiput(257.59,256.00)(0.477,2.046){7}{\rule{0.115pt}{1.620pt}}
\multiput(256.17,256.00)(5.000,15.638){2}{\rule{0.400pt}{0.810pt}}
\multiput(262.59,275.00)(0.477,1.935){7}{\rule{0.115pt}{1.540pt}}
\multiput(261.17,275.00)(5.000,14.804){2}{\rule{0.400pt}{0.770pt}}
\multiput(267.59,293.00)(0.482,1.485){9}{\rule{0.116pt}{1.233pt}}
\multiput(266.17,293.00)(6.000,14.440){2}{\rule{0.400pt}{0.617pt}}
\multiput(273.59,310.00)(0.477,1.823){7}{\rule{0.115pt}{1.460pt}}
\multiput(272.17,310.00)(5.000,13.970){2}{\rule{0.400pt}{0.730pt}}
\multiput(278.59,327.00)(0.477,1.601){7}{\rule{0.115pt}{1.300pt}}
\multiput(277.17,327.00)(5.000,12.302){2}{\rule{0.400pt}{0.650pt}}
\multiput(283.59,342.00)(0.482,1.304){9}{\rule{0.116pt}{1.100pt}}
\multiput(282.17,342.00)(6.000,12.717){2}{\rule{0.400pt}{0.550pt}}
\multiput(289.59,357.00)(0.477,1.489){7}{\rule{0.115pt}{1.220pt}}
\multiput(288.17,357.00)(5.000,11.468){2}{\rule{0.400pt}{0.610pt}}
\multiput(294.59,371.00)(0.477,1.378){7}{\rule{0.115pt}{1.140pt}}
\multiput(293.17,371.00)(5.000,10.634){2}{\rule{0.400pt}{0.570pt}}
\multiput(299.59,384.00)(0.477,1.267){7}{\rule{0.115pt}{1.060pt}}
\multiput(298.17,384.00)(5.000,9.800){2}{\rule{0.400pt}{0.530pt}}
\multiput(304.59,396.00)(0.482,0.852){9}{\rule{0.116pt}{0.767pt}}
\multiput(303.17,396.00)(6.000,8.409){2}{\rule{0.400pt}{0.383pt}}
\multiput(310.59,406.00)(0.477,1.044){7}{\rule{0.115pt}{0.900pt}}
\multiput(309.17,406.00)(5.000,8.132){2}{\rule{0.400pt}{0.450pt}}
\multiput(315.59,416.00)(0.477,0.933){7}{\rule{0.115pt}{0.820pt}}
\multiput(314.17,416.00)(5.000,7.298){2}{\rule{0.400pt}{0.410pt}}
\multiput(320.59,425.00)(0.477,0.710){7}{\rule{0.115pt}{0.660pt}}
\multiput(319.17,425.00)(5.000,5.630){2}{\rule{0.400pt}{0.330pt}}
\multiput(325.00,432.59)(0.491,0.482){9}{\rule{0.500pt}{0.116pt}}
\multiput(325.00,431.17)(4.962,6.000){2}{\rule{0.250pt}{0.400pt}}
\multiput(331.59,438.00)(0.477,0.599){7}{\rule{0.115pt}{0.580pt}}
\multiput(330.17,438.00)(5.000,4.796){2}{\rule{0.400pt}{0.290pt}}
\multiput(336.00,444.60)(0.627,0.468){5}{\rule{0.600pt}{0.113pt}}
\multiput(336.00,443.17)(3.755,4.000){2}{\rule{0.300pt}{0.400pt}}
\multiput(341.00,448.61)(0.909,0.447){3}{\rule{0.767pt}{0.108pt}}
\multiput(341.00,447.17)(3.409,3.000){2}{\rule{0.383pt}{0.400pt}}
\put(346,451.17){\rule{1.300pt}{0.400pt}}
\multiput(346.00,450.17)(3.302,2.000){2}{\rule{0.650pt}{0.400pt}}
\put(352,452.67){\rule{1.204pt}{0.400pt}}
\multiput(352.00,452.17)(2.500,1.000){2}{\rule{0.602pt}{0.400pt}}
\put(368,452.17){\rule{1.100pt}{0.400pt}}
\multiput(368.00,453.17)(2.717,-2.000){2}{\rule{0.550pt}{0.400pt}}
\multiput(373.00,450.95)(0.909,-0.447){3}{\rule{0.767pt}{0.108pt}}
\multiput(373.00,451.17)(3.409,-3.000){2}{\rule{0.383pt}{0.400pt}}
\multiput(378.00,447.95)(0.909,-0.447){3}{\rule{0.767pt}{0.108pt}}
\multiput(378.00,448.17)(3.409,-3.000){2}{\rule{0.383pt}{0.400pt}}
\multiput(383.00,444.93)(0.599,-0.477){7}{\rule{0.580pt}{0.115pt}}
\multiput(383.00,445.17)(4.796,-5.000){2}{\rule{0.290pt}{0.400pt}}
\multiput(389.00,439.93)(0.487,-0.477){7}{\rule{0.500pt}{0.115pt}}
\multiput(389.00,440.17)(3.962,-5.000){2}{\rule{0.250pt}{0.400pt}}
\multiput(394.00,434.93)(0.487,-0.477){7}{\rule{0.500pt}{0.115pt}}
\multiput(394.00,435.17)(3.962,-5.000){2}{\rule{0.250pt}{0.400pt}}
\multiput(399.59,428.59)(0.477,-0.599){7}{\rule{0.115pt}{0.580pt}}
\multiput(398.17,429.80)(5.000,-4.796){2}{\rule{0.400pt}{0.290pt}}
\multiput(404.59,422.65)(0.482,-0.581){9}{\rule{0.116pt}{0.567pt}}
\multiput(403.17,423.82)(6.000,-5.824){2}{\rule{0.400pt}{0.283pt}}
\multiput(410.59,415.26)(0.477,-0.710){7}{\rule{0.115pt}{0.660pt}}
\multiput(409.17,416.63)(5.000,-5.630){2}{\rule{0.400pt}{0.330pt}}
\multiput(415.59,407.93)(0.477,-0.821){7}{\rule{0.115pt}{0.740pt}}
\multiput(414.17,409.46)(5.000,-6.464){2}{\rule{0.400pt}{0.370pt}}
\multiput(420.59,400.37)(0.482,-0.671){9}{\rule{0.116pt}{0.633pt}}
\multiput(419.17,401.69)(6.000,-6.685){2}{\rule{0.400pt}{0.317pt}}
\multiput(426.59,391.60)(0.477,-0.933){7}{\rule{0.115pt}{0.820pt}}
\multiput(425.17,393.30)(5.000,-7.298){2}{\rule{0.400pt}{0.410pt}}
\multiput(431.59,382.60)(0.477,-0.933){7}{\rule{0.115pt}{0.820pt}}
\multiput(430.17,384.30)(5.000,-7.298){2}{\rule{0.400pt}{0.410pt}}
\multiput(436.59,373.60)(0.477,-0.933){7}{\rule{0.115pt}{0.820pt}}
\multiput(435.17,375.30)(5.000,-7.298){2}{\rule{0.400pt}{0.410pt}}
\multiput(441.59,365.09)(0.482,-0.762){9}{\rule{0.116pt}{0.700pt}}
\multiput(440.17,366.55)(6.000,-7.547){2}{\rule{0.400pt}{0.350pt}}
\multiput(447.59,355.60)(0.477,-0.933){7}{\rule{0.115pt}{0.820pt}}
\multiput(446.17,357.30)(5.000,-7.298){2}{\rule{0.400pt}{0.410pt}}
\multiput(452.59,346.60)(0.477,-0.933){7}{\rule{0.115pt}{0.820pt}}
\multiput(451.17,348.30)(5.000,-7.298){2}{\rule{0.400pt}{0.410pt}}
\multiput(457.59,337.26)(0.477,-1.044){7}{\rule{0.115pt}{0.900pt}}
\multiput(456.17,339.13)(5.000,-8.132){2}{\rule{0.400pt}{0.450pt}}
\multiput(462.59,328.09)(0.482,-0.762){9}{\rule{0.116pt}{0.700pt}}
\multiput(461.17,329.55)(6.000,-7.547){2}{\rule{0.400pt}{0.350pt}}
\multiput(468.59,318.26)(0.477,-1.044){7}{\rule{0.115pt}{0.900pt}}
\multiput(467.17,320.13)(5.000,-8.132){2}{\rule{0.400pt}{0.450pt}}
\multiput(473.59,308.60)(0.477,-0.933){7}{\rule{0.115pt}{0.820pt}}
\multiput(472.17,310.30)(5.000,-7.298){2}{\rule{0.400pt}{0.410pt}}
\multiput(478.59,299.60)(0.477,-0.933){7}{\rule{0.115pt}{0.820pt}}
\multiput(477.17,301.30)(5.000,-7.298){2}{\rule{0.400pt}{0.410pt}}
\multiput(483.59,291.09)(0.482,-0.762){9}{\rule{0.116pt}{0.700pt}}
\multiput(482.17,292.55)(6.000,-7.547){2}{\rule{0.400pt}{0.350pt}}
\multiput(489.59,281.60)(0.477,-0.933){7}{\rule{0.115pt}{0.820pt}}
\multiput(488.17,283.30)(5.000,-7.298){2}{\rule{0.400pt}{0.410pt}}
\multiput(494.59,272.60)(0.477,-0.933){7}{\rule{0.115pt}{0.820pt}}
\multiput(493.17,274.30)(5.000,-7.298){2}{\rule{0.400pt}{0.410pt}}
\multiput(499.59,264.37)(0.482,-0.671){9}{\rule{0.116pt}{0.633pt}}
\multiput(498.17,265.69)(6.000,-6.685){2}{\rule{0.400pt}{0.317pt}}
\multiput(505.59,255.60)(0.477,-0.933){7}{\rule{0.115pt}{0.820pt}}
\multiput(504.17,257.30)(5.000,-7.298){2}{\rule{0.400pt}{0.410pt}}
\multiput(510.59,246.93)(0.477,-0.821){7}{\rule{0.115pt}{0.740pt}}
\multiput(509.17,248.46)(5.000,-6.464){2}{\rule{0.400pt}{0.370pt}}
\multiput(515.59,238.93)(0.477,-0.821){7}{\rule{0.115pt}{0.740pt}}
\multiput(514.17,240.46)(5.000,-6.464){2}{\rule{0.400pt}{0.370pt}}
\multiput(520.59,231.65)(0.482,-0.581){9}{\rule{0.116pt}{0.567pt}}
\multiput(519.17,232.82)(6.000,-5.824){2}{\rule{0.400pt}{0.283pt}}
\multiput(526.59,224.26)(0.477,-0.710){7}{\rule{0.115pt}{0.660pt}}
\multiput(525.17,225.63)(5.000,-5.630){2}{\rule{0.400pt}{0.330pt}}
\multiput(531.59,217.26)(0.477,-0.710){7}{\rule{0.115pt}{0.660pt}}
\multiput(530.17,218.63)(5.000,-5.630){2}{\rule{0.400pt}{0.330pt}}
\multiput(536.59,210.26)(0.477,-0.710){7}{\rule{0.115pt}{0.660pt}}
\multiput(535.17,211.63)(5.000,-5.630){2}{\rule{0.400pt}{0.330pt}}
\multiput(541.00,204.93)(0.491,-0.482){9}{\rule{0.500pt}{0.116pt}}
\multiput(541.00,205.17)(4.962,-6.000){2}{\rule{0.250pt}{0.400pt}}
\multiput(547.59,197.59)(0.477,-0.599){7}{\rule{0.115pt}{0.580pt}}
\multiput(546.17,198.80)(5.000,-4.796){2}{\rule{0.400pt}{0.290pt}}
\multiput(552.59,191.59)(0.477,-0.599){7}{\rule{0.115pt}{0.580pt}}
\multiput(551.17,192.80)(5.000,-4.796){2}{\rule{0.400pt}{0.290pt}}
\multiput(557.00,186.93)(0.491,-0.482){9}{\rule{0.500pt}{0.116pt}}
\multiput(557.00,187.17)(4.962,-6.000){2}{\rule{0.250pt}{0.400pt}}
\multiput(563.00,180.93)(0.487,-0.477){7}{\rule{0.500pt}{0.115pt}}
\multiput(563.00,181.17)(3.962,-5.000){2}{\rule{0.250pt}{0.400pt}}
\multiput(568.00,175.93)(0.487,-0.477){7}{\rule{0.500pt}{0.115pt}}
\multiput(568.00,176.17)(3.962,-5.000){2}{\rule{0.250pt}{0.400pt}}
\multiput(573.00,170.94)(0.627,-0.468){5}{\rule{0.600pt}{0.113pt}}
\multiput(573.00,171.17)(3.755,-4.000){2}{\rule{0.300pt}{0.400pt}}
\multiput(578.00,166.93)(0.599,-0.477){7}{\rule{0.580pt}{0.115pt}}
\multiput(578.00,167.17)(4.796,-5.000){2}{\rule{0.290pt}{0.400pt}}
\multiput(584.00,161.94)(0.627,-0.468){5}{\rule{0.600pt}{0.113pt}}
\multiput(584.00,162.17)(3.755,-4.000){2}{\rule{0.300pt}{0.400pt}}
\multiput(589.00,157.95)(0.909,-0.447){3}{\rule{0.767pt}{0.108pt}}
\multiput(589.00,158.17)(3.409,-3.000){2}{\rule{0.383pt}{0.400pt}}
\multiput(594.00,154.94)(0.627,-0.468){5}{\rule{0.600pt}{0.113pt}}
\multiput(594.00,155.17)(3.755,-4.000){2}{\rule{0.300pt}{0.400pt}}
\multiput(599.00,150.95)(1.132,-0.447){3}{\rule{0.900pt}{0.108pt}}
\multiput(599.00,151.17)(4.132,-3.000){2}{\rule{0.450pt}{0.400pt}}
\multiput(605.00,147.95)(0.909,-0.447){3}{\rule{0.767pt}{0.108pt}}
\multiput(605.00,148.17)(3.409,-3.000){2}{\rule{0.383pt}{0.400pt}}
\multiput(610.00,144.95)(0.909,-0.447){3}{\rule{0.767pt}{0.108pt}}
\multiput(610.00,145.17)(3.409,-3.000){2}{\rule{0.383pt}{0.400pt}}
\multiput(615.00,141.95)(1.132,-0.447){3}{\rule{0.900pt}{0.108pt}}
\multiput(615.00,142.17)(4.132,-3.000){2}{\rule{0.450pt}{0.400pt}}
\put(621,138.17){\rule{1.100pt}{0.400pt}}
\multiput(621.00,139.17)(2.717,-2.000){2}{\rule{0.550pt}{0.400pt}}
\multiput(626.00,136.95)(0.909,-0.447){3}{\rule{0.767pt}{0.108pt}}
\multiput(626.00,137.17)(3.409,-3.000){2}{\rule{0.383pt}{0.400pt}}
\put(631,133.17){\rule{1.100pt}{0.400pt}}
\multiput(631.00,134.17)(2.717,-2.000){2}{\rule{0.550pt}{0.400pt}}
\put(636,131.17){\rule{1.300pt}{0.400pt}}
\multiput(636.00,132.17)(3.302,-2.000){2}{\rule{0.650pt}{0.400pt}}
\put(642,129.67){\rule{1.204pt}{0.400pt}}
\multiput(642.00,130.17)(2.500,-1.000){2}{\rule{0.602pt}{0.400pt}}
\put(647,128.17){\rule{1.100pt}{0.400pt}}
\multiput(647.00,129.17)(2.717,-2.000){2}{\rule{0.550pt}{0.400pt}}
\put(652,126.17){\rule{1.100pt}{0.400pt}}
\multiput(652.00,127.17)(2.717,-2.000){2}{\rule{0.550pt}{0.400pt}}
\put(657,124.67){\rule{1.445pt}{0.400pt}}
\multiput(657.00,125.17)(3.000,-1.000){2}{\rule{0.723pt}{0.400pt}}
\put(663,123.67){\rule{1.204pt}{0.400pt}}
\multiput(663.00,124.17)(2.500,-1.000){2}{\rule{0.602pt}{0.400pt}}
\put(668,122.67){\rule{1.204pt}{0.400pt}}
\multiput(668.00,123.17)(2.500,-1.000){2}{\rule{0.602pt}{0.400pt}}
\put(673,121.67){\rule{1.204pt}{0.400pt}}
\multiput(673.00,122.17)(2.500,-1.000){2}{\rule{0.602pt}{0.400pt}}
\put(678,120.67){\rule{1.445pt}{0.400pt}}
\multiput(678.00,121.17)(3.000,-1.000){2}{\rule{0.723pt}{0.400pt}}
\put(684,119.67){\rule{1.204pt}{0.400pt}}
\multiput(684.00,120.17)(2.500,-1.000){2}{\rule{0.602pt}{0.400pt}}
\put(689,118.67){\rule{1.204pt}{0.400pt}}
\multiput(689.00,119.17)(2.500,-1.000){2}{\rule{0.602pt}{0.400pt}}
\put(357.0,454.0){\rule[-0.200pt]{2.650pt}{0.400pt}}
\put(700,117.67){\rule{1.204pt}{0.400pt}}
\multiput(700.00,118.17)(2.500,-1.000){2}{\rule{0.602pt}{0.400pt}}
\put(705,116.67){\rule{1.204pt}{0.400pt}}
\multiput(705.00,117.17)(2.500,-1.000){2}{\rule{0.602pt}{0.400pt}}
\put(694.0,119.0){\rule[-0.200pt]{1.445pt}{0.400pt}}
\put(715,115.67){\rule{1.445pt}{0.400pt}}
\multiput(715.00,116.17)(3.000,-1.000){2}{\rule{0.723pt}{0.400pt}}
\put(710.0,117.0){\rule[-0.200pt]{1.204pt}{0.400pt}}
\put(731,114.67){\rule{1.204pt}{0.400pt}}
\multiput(731.00,115.17)(2.500,-1.000){2}{\rule{0.602pt}{0.400pt}}
\put(721.0,116.0){\rule[-0.200pt]{2.409pt}{0.400pt}}
\put(736.0,115.0){\rule[-0.200pt]{2.650pt}{0.400pt}}
\put(659,562){\makebox(0,0)[r]{$v_{\theta}^{-1}=30$}}
\put(681.0,562.0){\rule[-0.200pt]{15.899pt}{0.400pt}}
\put(220.0,113.0){\rule[-0.200pt]{0.400pt}{24.813pt}}
\put(220.0,216.0){\rule[-0.200pt]{4.336pt}{0.400pt}}
\put(238.0,113.0){\rule[-0.200pt]{0.400pt}{24.813pt}}
\put(220.0,113.0){\rule[-0.200pt]{4.336pt}{0.400pt}}
\put(238.0,113.0){\rule[-0.200pt]{0.400pt}{129.845pt}}
\put(238.0,652.0){\rule[-0.200pt]{4.095pt}{0.400pt}}
\put(255.0,113.0){\rule[-0.200pt]{0.400pt}{129.845pt}}
\put(238.0,113.0){\rule[-0.200pt]{4.095pt}{0.400pt}}
\put(255.0,113.0){\rule[-0.200pt]{0.400pt}{116.596pt}}
\put(255.0,597.0){\rule[-0.200pt]{4.336pt}{0.400pt}}
\put(273.0,113.0){\rule[-0.200pt]{0.400pt}{116.596pt}}
\put(255.0,113.0){\rule[-0.200pt]{4.336pt}{0.400pt}}
\put(273.0,113.0){\rule[-0.200pt]{0.400pt}{86.483pt}}
\put(273.0,472.0){\rule[-0.200pt]{4.095pt}{0.400pt}}
\put(290.0,113.0){\rule[-0.200pt]{0.400pt}{86.483pt}}
\put(273.0,113.0){\rule[-0.200pt]{4.095pt}{0.400pt}}
\put(290.0,113.0){\rule[-0.200pt]{0.400pt}{72.511pt}}
\put(290.0,414.0){\rule[-0.200pt]{4.336pt}{0.400pt}}
\put(308.0,113.0){\rule[-0.200pt]{0.400pt}{72.511pt}}
\put(290.0,113.0){\rule[-0.200pt]{4.336pt}{0.400pt}}
\put(308.0,113.0){\rule[-0.200pt]{0.400pt}{52.275pt}}
\put(308.0,330.0){\rule[-0.200pt]{4.095pt}{0.400pt}}
\put(325.0,113.0){\rule[-0.200pt]{0.400pt}{52.275pt}}
\put(308.0,113.0){\rule[-0.200pt]{4.095pt}{0.400pt}}
\put(325.0,113.0){\rule[-0.200pt]{0.400pt}{58.539pt}}
\put(325.0,356.0){\rule[-0.200pt]{4.336pt}{0.400pt}}
\put(343.0,113.0){\rule[-0.200pt]{0.400pt}{58.539pt}}
\put(325.0,113.0){\rule[-0.200pt]{4.336pt}{0.400pt}}
\put(343.0,113.0){\rule[-0.200pt]{0.400pt}{51.553pt}}
\put(343.0,327.0){\rule[-0.200pt]{4.336pt}{0.400pt}}
\put(361.0,113.0){\rule[-0.200pt]{0.400pt}{51.553pt}}
\put(343.0,113.0){\rule[-0.200pt]{4.336pt}{0.400pt}}
\put(361.0,113.0){\rule[-0.200pt]{0.400pt}{56.611pt}}
\put(361.0,348.0){\rule[-0.200pt]{4.095pt}{0.400pt}}
\put(378.0,113.0){\rule[-0.200pt]{0.400pt}{56.611pt}}
\put(361.0,113.0){\rule[-0.200pt]{4.095pt}{0.400pt}}
\put(378.0,113.0){\rule[-0.200pt]{0.400pt}{45.771pt}}
\put(378.0,303.0){\rule[-0.200pt]{4.336pt}{0.400pt}}
\put(396.0,113.0){\rule[-0.200pt]{0.400pt}{45.771pt}}
\put(378.0,113.0){\rule[-0.200pt]{4.336pt}{0.400pt}}
\put(396.0,113.0){\rule[-0.200pt]{0.400pt}{28.185pt}}
\put(396.0,230.0){\rule[-0.200pt]{4.095pt}{0.400pt}}
\put(413.0,113.0){\rule[-0.200pt]{0.400pt}{28.185pt}}
\put(396.0,113.0){\rule[-0.200pt]{4.095pt}{0.400pt}}
\put(413.0,113.0){\rule[-0.200pt]{0.400pt}{29.872pt}}
\put(413.0,237.0){\rule[-0.200pt]{4.336pt}{0.400pt}}
\put(431.0,113.0){\rule[-0.200pt]{0.400pt}{29.872pt}}
\put(413.0,113.0){\rule[-0.200pt]{4.336pt}{0.400pt}}
\put(431.0,113.0){\rule[-0.200pt]{0.400pt}{29.872pt}}
\put(431.0,237.0){\rule[-0.200pt]{4.095pt}{0.400pt}}
\put(448.0,113.0){\rule[-0.200pt]{0.400pt}{29.872pt}}
\put(431.0,113.0){\rule[-0.200pt]{4.095pt}{0.400pt}}
\put(448.0,113.0){\rule[-0.200pt]{0.400pt}{27.703pt}}
\put(448.0,228.0){\rule[-0.200pt]{4.336pt}{0.400pt}}
\put(466.0,113.0){\rule[-0.200pt]{0.400pt}{27.703pt}}
\put(448.0,113.0){\rule[-0.200pt]{4.336pt}{0.400pt}}
\put(466.0,113.0){\rule[-0.200pt]{0.400pt}{26.981pt}}
\put(466.0,225.0){\rule[-0.200pt]{4.095pt}{0.400pt}}
\put(483.0,113.0){\rule[-0.200pt]{0.400pt}{26.981pt}}
\put(466.0,113.0){\rule[-0.200pt]{4.095pt}{0.400pt}}
\put(483.0,113.0){\rule[-0.200pt]{0.400pt}{25.294pt}}
\put(483.0,218.0){\rule[-0.200pt]{4.336pt}{0.400pt}}
\put(501.0,113.0){\rule[-0.200pt]{0.400pt}{25.294pt}}
\put(483.0,113.0){\rule[-0.200pt]{4.336pt}{0.400pt}}
\put(501.0,113.0){\rule[-0.200pt]{0.400pt}{24.090pt}}
\put(501.0,213.0){\rule[-0.200pt]{4.336pt}{0.400pt}}
\put(519.0,113.0){\rule[-0.200pt]{0.400pt}{24.090pt}}
\put(501.0,113.0){\rule[-0.200pt]{4.336pt}{0.400pt}}
\put(519.0,113.0){\rule[-0.200pt]{0.400pt}{21.922pt}}
\put(519.0,204.0){\rule[-0.200pt]{4.095pt}{0.400pt}}
\put(536.0,113.0){\rule[-0.200pt]{0.400pt}{21.922pt}}
\put(519.0,113.0){\rule[-0.200pt]{4.095pt}{0.400pt}}
\put(536.0,113.0){\rule[-0.200pt]{0.400pt}{12.527pt}}
\put(536.0,165.0){\rule[-0.200pt]{4.336pt}{0.400pt}}
\put(554.0,113.0){\rule[-0.200pt]{0.400pt}{12.527pt}}
\put(536.0,113.0){\rule[-0.200pt]{4.336pt}{0.400pt}}
\put(554.0,113.0){\rule[-0.200pt]{0.400pt}{15.418pt}}
\put(554.0,177.0){\rule[-0.200pt]{4.095pt}{0.400pt}}
\put(571.0,113.0){\rule[-0.200pt]{0.400pt}{15.418pt}}
\put(554.0,113.0){\rule[-0.200pt]{4.095pt}{0.400pt}}
\put(571.0,113.0){\rule[-0.200pt]{0.400pt}{16.622pt}}
\put(571.0,182.0){\rule[-0.200pt]{4.336pt}{0.400pt}}
\put(589.0,113.0){\rule[-0.200pt]{0.400pt}{16.622pt}}
\put(571.0,113.0){\rule[-0.200pt]{4.336pt}{0.400pt}}
\put(589.0,113.0){\rule[-0.200pt]{0.400pt}{14.454pt}}
\put(589.0,173.0){\rule[-0.200pt]{4.095pt}{0.400pt}}
\put(606.0,113.0){\rule[-0.200pt]{0.400pt}{14.454pt}}
\put(589.0,113.0){\rule[-0.200pt]{4.095pt}{0.400pt}}
\put(606.0,113.0){\rule[-0.200pt]{0.400pt}{8.913pt}}
\put(606.0,150.0){\rule[-0.200pt]{4.336pt}{0.400pt}}
\put(624.0,113.0){\rule[-0.200pt]{0.400pt}{8.913pt}}
\put(606.0,113.0){\rule[-0.200pt]{4.336pt}{0.400pt}}
\put(624.0,113.0){\rule[-0.200pt]{0.400pt}{11.804pt}}
\put(624.0,162.0){\rule[-0.200pt]{4.336pt}{0.400pt}}
\put(642.0,113.0){\rule[-0.200pt]{0.400pt}{11.804pt}}
\put(624.0,113.0){\rule[-0.200pt]{4.336pt}{0.400pt}}
\put(642.0,113.0){\rule[-0.200pt]{0.400pt}{7.950pt}}
\put(642.0,146.0){\rule[-0.200pt]{4.095pt}{0.400pt}}
\put(659.0,113.0){\rule[-0.200pt]{0.400pt}{7.950pt}}
\put(642.0,113.0){\rule[-0.200pt]{4.095pt}{0.400pt}}
\put(659.0,113.0){\rule[-0.200pt]{0.400pt}{8.913pt}}
\put(659.0,150.0){\rule[-0.200pt]{4.336pt}{0.400pt}}
\put(677.0,113.0){\rule[-0.200pt]{0.400pt}{8.913pt}}
\put(659.0,113.0){\rule[-0.200pt]{4.336pt}{0.400pt}}
\put(677.0,113.0){\rule[-0.200pt]{0.400pt}{6.745pt}}
\put(677.0,141.0){\rule[-0.200pt]{4.095pt}{0.400pt}}
\put(694.0,113.0){\rule[-0.200pt]{0.400pt}{6.745pt}}
\put(677.0,113.0){\rule[-0.200pt]{4.095pt}{0.400pt}}
\put(694.0,113.0){\rule[-0.200pt]{0.400pt}{7.227pt}}
\put(694.0,143.0){\rule[-0.200pt]{4.336pt}{0.400pt}}
\put(712.0,113.0){\rule[-0.200pt]{0.400pt}{7.227pt}}
\put(694.0,113.0){\rule[-0.200pt]{4.336pt}{0.400pt}}
\put(712.0,113.0){\rule[-0.200pt]{0.400pt}{4.577pt}}
\put(712.0,132.0){\rule[-0.200pt]{4.095pt}{0.400pt}}
\put(729.0,113.0){\rule[-0.200pt]{0.400pt}{4.577pt}}
\put(712.0,113.0){\rule[-0.200pt]{4.095pt}{0.400pt}}
\put(729.0,113.0){\rule[-0.200pt]{0.400pt}{3.132pt}}
\put(729.0,126.0){\rule[-0.200pt]{4.336pt}{0.400pt}}
\put(747.0,113.0){\rule[-0.200pt]{0.400pt}{3.132pt}}
\put(729.0,113.0){\rule[-0.200pt]{4.336pt}{0.400pt}}
\end{picture}